\documentclass[twoside,journey]{IEEEtran}
\usepackage{makecell}
\usepackage{hyperref}
\usepackage{array}
\usepackage{graphicx,amssymb,amsmath}
\usepackage{multicol}
\usepackage[noadjust]{cite}
\usepackage{setspace}
\usepackage{subfigure}
\usepackage{graphicx}
\usepackage{float}
\usepackage{url}
\usepackage{algorithm}
\usepackage{algpseudocode}

\usepackage{stfloats}
\usepackage{amsthm,pifont}
\usepackage{flushend}
\usepackage{cases,subeqnarray}
\usepackage{bm,multirow,bigstrut}
\usepackage{amsmath, amsthm, amssymb}
\usepackage{textcomp}
\usepackage{latexsym,bm}
\usepackage{booktabs}
\usepackage{xcolor}
\usepackage{mathtools}
\usepackage{dsfont}
\usepackage{extarrows}
\usepackage{epsfig}
\usepackage{epsfig}
\usepackage{epstopdf}
\usepackage{algorithmicx}
\theoremstyle{plain}

\theoremstyle{plain}
\newtheorem{rem}{Remark}
\newtheorem{them}{Theorem}

\usepackage{amsmath}

\newtheorem{lemma}{Lemma}
\IEEEoverridecommandlockouts
\begin{document}
%----------------------------title&author&thanks----------------------------
\title{AI-Generated Incentive Mechanism and Full-Duplex Semantic Communications for Information Sharing}
%Reconfigurable Intelligent Surface-Aided RIS-Aided Data Compression For 6G
\author{Hongyang~Du, Jiacheng~Wang, Dusit~Niyato,~\IEEEmembership{Fellow,~IEEE}, Jiawen~Kang, Zehui Xiong, and Dong~In~Kim,~\IEEEmembership{Fellow,~IEEE}
\thanks{H.~Du, J.~Wang, and D. Niyato are with the School of Computer Science and Engineering, Nanyang Technological University, Singapore (e-mail: hongyang001@e.ntu.edu.sg, dniyato@ntu.edu.sg, jcwang\_cq@foxmail.com)}
\thanks{J. Kang is with the School of Automation, Guangdong University of Technology, China. (e-mail: kavinkang@gdut.edu.cn)}
\thanks{Z. Xiong is with the Pillar of Information Systems Technology and Design, Singapore University of Technology and Design, Singapore (e-mail: zehui\_xiong@sutd.edu.sg)}
\thanks{D. I. Kim is with the Department of Electrical and Computer Engineering, Sungkyunkwan University, South Korea (e-mail: dikim@skku.ac.kr)}
}
\maketitle
%----------------------------abstract----------------------------
\vspace{-1cm}
\begin{abstract}
The next generation of Internet services, such as Metaverse, rely on mixed reality (MR) technology to provide immersive user experiences. However, limited computation power of MR headset-mounted devices (HMDs) hinders the deployment of such services. Therefore, we propose an efficient information-sharing scheme based on full-duplex device-to-device (D2D) semantic communications to address this issue. Our approach enables users to avoid heavy and repetitive computational tasks, such as artificial intelligence-generated content (AIGC) in the view images of all MR users. Specifically, a user can transmit the generated content and semantic information extracted from their view image to nearby users, who can then use this information to obtain the spatial matching of computation results under their view images. We analyze the performance of full-duplex D2D communications, including the achievable rate and bit error probability, by using generalized small-scale fading models. To facilitate semantic information sharing among users, we design a contract theoretic AI-generated incentive mechanism. The proposed diffusion model generates the optimal contract design, outperforming two deep reinforcement learning algorithms, i.e., proximal policy optimization and soft actor-critic algorithms. Our numerical analysis experiment proves the effectiveness of our proposed methods. The code for this paper is available at \url{https://github.com/HongyangDu/SemSharing}.
\end{abstract}
%----------------------------keywords----------------------------
%\newpage
\IEEEpeerreviewmaketitle
%----------------------------introduction----------------------------
\section{Introduction}
The development of computer science and wireless networks has continuously enriched the ways of interaction, social communications, and transactions among users, thus playing an irreplaceable role in daily life~\cite{li2020esync}. From the users' perspective, three significant waves of technological innovation have been witnessed, i.e., the introduction of personal computers, the Internet, and mobile devices. The fourth wave is currently unfolding based on immersive technologies, such as mixed reality (MR), with the goal of forming a ubiquitous virtual space, i.e., Metaverse~\cite{wang2022survey}. Defined as a collective virtual shared world of value co-creation, Metaverse, which is parallel to the physical world, incorporates different technologies to enhance physical spaces, products, and services, thereby revolutionizing online education, business, remote work, and entertainment.

\begin{figure}[t]
\centering
\includegraphics[width=0.4\textwidth]{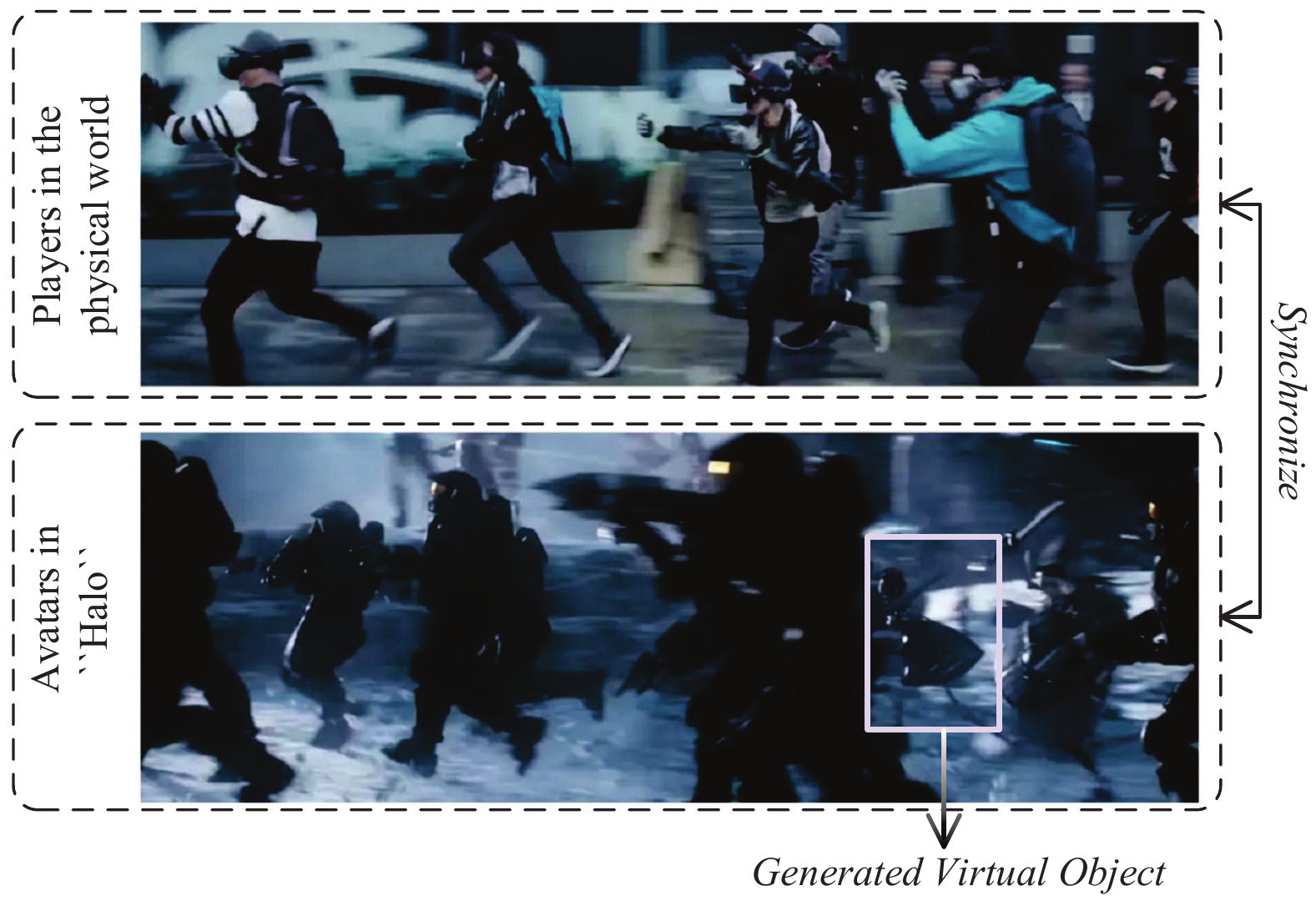}%
\caption{In ``Ready Player One'', MR-headset-wearing players run through the streets with controllers in hand. However, they look like characters from the game ``Halo'' running into battle in the Oasis.}
\label{ready}
\end{figure}

Immersive technologies, which benefit from powerful computing and rendering capacities, provide users with a natural augmentation of the real world, allowing individuals to generate avatars and seamlessly experience different services. For instance, users can wear MR equipment such as headset-mounted devices (HMDs) and hold controllers to play virtual battle games, just like the scene in the movie ``Ready Player One'' as shown in Fig.~\ref{ready}. During the game, the user's position in the physical world and the environmental information of the surrounding physical space are collected by devices such as cameras and transmitted to the Metaverse service provider (MSP). The MSP leverages the obtained information to build a virtual gunfight scene and display it to users. Simultaneously, the MR equipment collects the data that describes the user's motion and posture in real time and updates it to the MSP.
Through a similar way, some other applications, such as virtual tourism or AIGC services~\cite{du2023enabling}, can also be offered in Metaverse. As stated in \cite{han2018user}, more than $84\%$ of consumers around the world are interested in using MR for travel experiences, and $42\%$ believe that augmented reality (AR) and virtual reality (VR) are the future of tourism. The reason is that MR enables users to have more engaged and diverse encounters. For instance, travelers in Metaverse can freely choose their destination or generate virtual objects without being restricted by physical limitations and rules. Furthermore, through human-computer interaction technologies, users can interact with other virtual travelers, tour guides, or even AIGC in Metaverse to achieve an immersive experience.

While the above applications are fascinating, several difficulties need to be addressed. {\textit{One of the most representative and urgent problems to be solved is that the computational tasks of MR Head-Mounted Displays (HMDs) are overly heavy.}} One reason is that, as each player is relatively independent, the scene displayed to the user through the MR HMD needs to be computed and rendered separately, which consumes a lot of computing and transmission resources. For instance, the virtual battle game depicted in Fig.~\ref{ready} exemplifies a scenario where subsequent players may be subjected to the same or similar in-game events as the previous players, leading to a repetition of computations and rendering virtual objects. One solution to this challenge is establishing an information-sharing scheme among users. Specifically, the view image, i.e., the real-world scene displayed on the MR HMD, serves as a reference point for detecting safe, walkable areas, also known as free-space information, based on the user's unique view image. By sharing the computation results, such as AIGC, and the view image with other users, the users can avoid repeating computations by comparing received view images to determine the free-space in their own view images and localizing the AIGC. While the information sharing scheme is a viable solution, transmitting such information among users requires a significant amount of transmission resources, particularly for high-definition view images.
% =======

In response to the problems above, this paper presents a full-duplex semantic communications framework that aims to enable an efficient information sharing among users. The proposed framework leverages a semantic coding algorithm to extract semantic features from the view image and then transmits them to other users using full-duplex communication. Users can use the received semantic features to obtain free-space information through lightweight semantic matching instead of performing repeated computations. This significantly reduces the computing resources consumed by each user. Furthermore, the full-duplex transmission of semantic information, which has a smaller data size than the original data, minimizes the transmission resource consumption of users.

Moreover, appropriate incentives must be established to motivate users to share semantic information efficiently and effectively~\cite{yang2022semantic,li2022data}. A straightforward and effective way to achieve this is to have the semantic information receiver (SIR) pay an incentive to the semantic information provider (SIP). Contract theory can be used to model the payment scheme for both parties, maximizing the utility of the SIR while meeting the incentive compatibility (IC) and individual rationality (IR) properties of both parties~\cite{zhang2017survey}. However, to solve the optimal contract, the recipient of the semantic information must know various characteristics of the sender, such as the cost per unit of transmit power. Additionally, different wireless transmission conditions, such as transmission distance, small-scale fading channels~\cite{durgin2002new}, and co-channel interference, can affect the optimal contract solution~\cite{du2022attention}. These multidimensional conditions make it challenging to solve the contract design problem with traditional mathematical methods. Although deep reinforcement learning (DRL)-based solutions can find the contract design without specific environmental parameters, the high state dimension can make reinforcement learning-based approaches challenging to solve or prone to falling into a local optimum~\cite{silver2016mastering,garnier2021review}. To address the aforementioned problems, taking inspiration from the recent remarkable achievements of generative artificial intelligence (AI) techniques in content generation, specifically diffusion models, we want to leverage generative AI to generate optimum contract designs. Our contributions are summarized as
\begin{itemize}
\item We propose a comprehensive framework for full-duplex device-to-device (D2D) semantic communication that enables efficient synchronization of free-space and semantic information among multiple MR users. Our proposed approach eliminates the need for individual users to compute the positions of virtual objects independently in their field of view.
\item Within the proposed framework, we consider the effects of wireless fading and self-interference issues that arise in D2D full-duplex communications, and analyzes their impact on the transmission of semantic information. We derive closed-form expressions for both the bit error probability (BEP) and the achievable rate, which are key performance metrics in evaluating the effectiveness of the full-duplex system.
\item To promote the sharing of semantic information among users, we present an incentive mechanism design based on contract theory. Specifically, we utilize generative AI, i.e., diffusion model, to design optimal contracts. To evaluate the effectiveness of the AI-generated incentive mechanism, we conduct the numerical analysis and show that our method outperforms the reinforcement learning-based scheme. {\textit{To the best of our knowledge, this is the first paper that uses diffusion model for incentive design.}}
\end{itemize}
The remainder of the paper is organized as follows: In Section \ref{S2}, we review the related work in the literature. In Section \ref{S3}, we introduce the semantic-aware free-space information sharing mechanism and propose an end-to-end full-duplex communication system model under generalized fading channel models. Section~\ref{S4} presents the semantic encoding and matching algorithms. In Section~\ref{S5}, we propose the diffusion model-based AI-generated contract algorithm. We then perform sound mathematical analysis of the key performance indicators in D2D wireless communications, i.e., achievable rate and BEP. Section~\ref{S6} presents the experiment results. Finally, Section \ref{S7} concludes this paper.

\section{Related Work}\label{S2}
In this section, we provide a brief review of three related techniques, i.e., full-duplex communications, visual simultaneous localization and mapping, and mixed reality-aided Metaverse services.
\subsection{Full-Duplex Communications}
The full-duplex communications technique allows users to transmit on the same band simultaneously while receiving \cite{alves2020full,kim2015survey}. Compared with half-duplex communications, full-duplex communications can further improve spectral efficiency, even doubling it if sufficient self-interference cancellation (SIC) is achieved~\cite{kolodziej2019band}. Motivated by this performance advantage, many research works have been done \cite{han2019power}. For the cellular networks covered by full-duplex D2D users with residual self-interference, the authors in \cite{ali2016modeling} developed a stochastic geometry-based tractable analytical paradigm to compute the outage probability and rate. The authors in \cite{liu2018capacity} analyzed the capacity improvement of full-duplex D2D underlying cellular networks, concluding that the capacity improvement of full-duplex D2D communications is more significant than that of half-duplex D2D communication if sufficient self-interference cancellation is achieved. Furthermore, the authors in \cite{lee2014power} and \cite{feng2013device} investigate full-duplex D2D communication from power control and resource allocation perspectives, respectively. However, work has yet to be done to analyze the performance metrics of full-duplex communication systems considering generalized fading channels and interference effects, i.e., self-interference and co-channel interference with signals of arbitrary paths.

\subsection{Visual Simultaneous Localization and Mapping}\label{slam}
The visual simultaneous localization and mapping (SLAM) technique uses data gathered by a camera-equipped platform to build a map of the area around the platform while simultaneously locating it inside the map. For the static environment or an environment with few dynamic elements, several visual SLAM systems have been proposed, such as MonoSLAM \cite{davison2007monoslam}, ORB-SLAM \cite{mur2015orb}, and DSO \cite{engel2017direct}. Specifically, MonoSLAM is believed to be the first filter-based approach that brings the SLAM problem from the robotic community into pure vision. For dynamic environments, researchers have also developed many systems \cite{xiao2019dynamic,wang2019computationally,xiao2019dynamic}. Authors in \cite{xiao2019dynamic} use the object detection network to obtain prior knowledge about dynamic objects and employ a selective tracking algorithm to process the features of dynamic objects, so as to reduce the pose estimation error. In \cite{wang2019computationally}, a step-by-step method is proposed, which consists of object detection and contour extraction, to extract semantic information of dynamic objects in a computationally more efficient way. With the rise of wearable devices and advances of matching and detection algorithms in computer vision, such like SuperPoint~\cite{detone2018superpoint} and SuperGlue~\cite{sarlin2020superglue}, the visual SLAM would benefit more applications, especially in Metaverse. Although promising, new problems, such as efficient and effective wireless information sharing among users, need to be solved urgently.

\subsection{Mixed Reality-Aided Metaverse}
MR is a general term for a type of technology, including but not limited to the well-known AR and VR, which can seamlessly integrate and merge 3D content into the user's physical environment \cite{rauschnabel2021augmented} and enables users to travel freely between the real world and Metaverse. With the support of MR, the MSP can provide users with more services, including educational services, interventional services, communication services, online entertainment, and so forth. The popular Pokemon G \cite{kerdvibulvech2022exploring} can be considered as a miniature of online game in Metaverse. The game uses GPS to locate the player in the physical world and displays creatures based on the obtained location as if they are in the player's real-world through the mobile device's camera. On this basis, players can capture, train, and battle the virtual creatures. Although interesting, these systems repeatedly compute and render similar or identical scenes for different players, resulting in a huge waste of computing resources, especially when a large number of players are playing at the same time. Moreover, when different users interact with each other, the transmission of information, particularly the sharing of high-definition images or video streams, also consumes a lot of resources. These problems need to be solved urgently before these systems are further extended into Metaverse.

Motivated by the aforementioned research gaps, we propose a semantic-aware free-space information sharing mechanism to reduce the network resources consumption. To further improve the communication efficiency, the D2D full-duplex techniques is used among users.

\section{System Model}\label{S3}
In this section, we present the semantic-aware free-space information sharing mechanism and study a D2D full-duplex communications system.
\subsection{Semantic-Aware Free-Space Information Sharing}\label{yuanli}
\begin{figure}[t]
\centering
\includegraphics[width=0.48\textwidth]{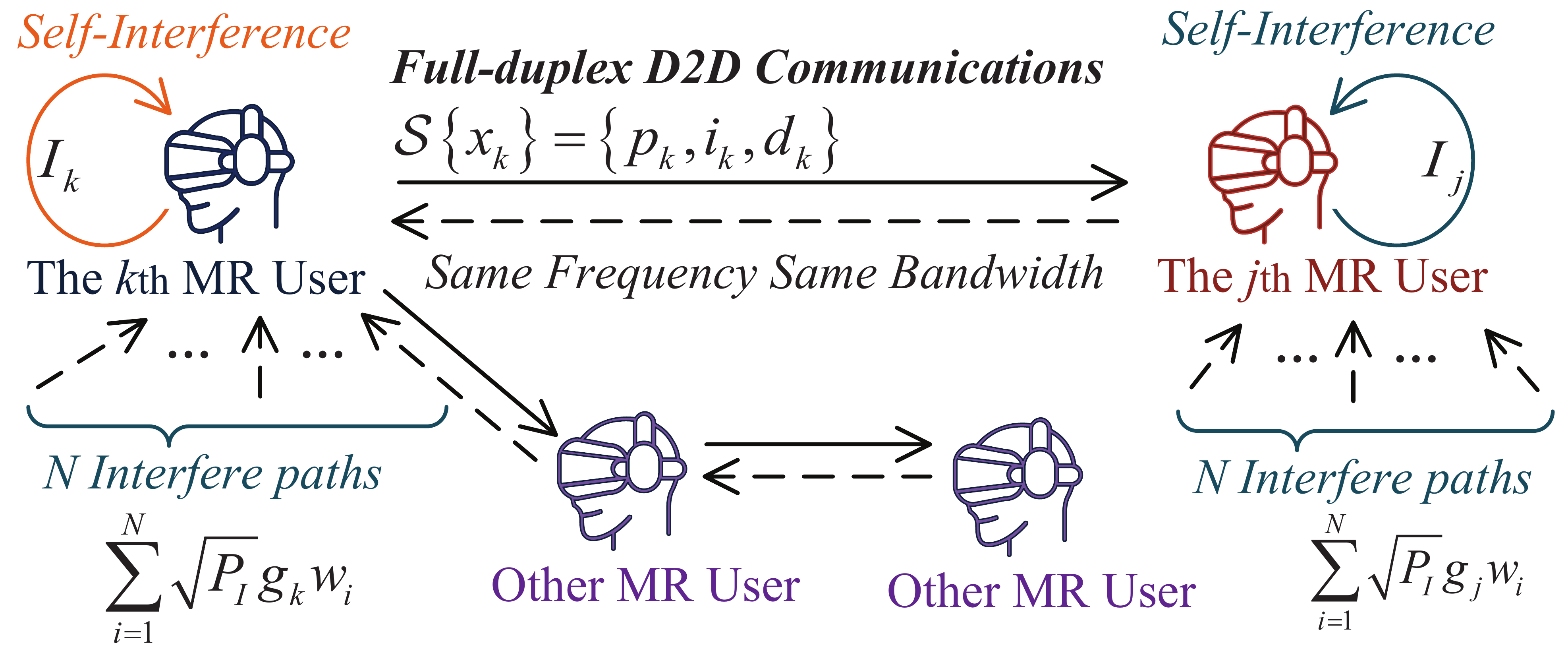}%
\caption{Device-to-device full-duplex wireless communication system model.}
\label{fwmodel}
\end{figure}
\begin{figure*}[t]
\centering
\includegraphics[width=0.96\textwidth]{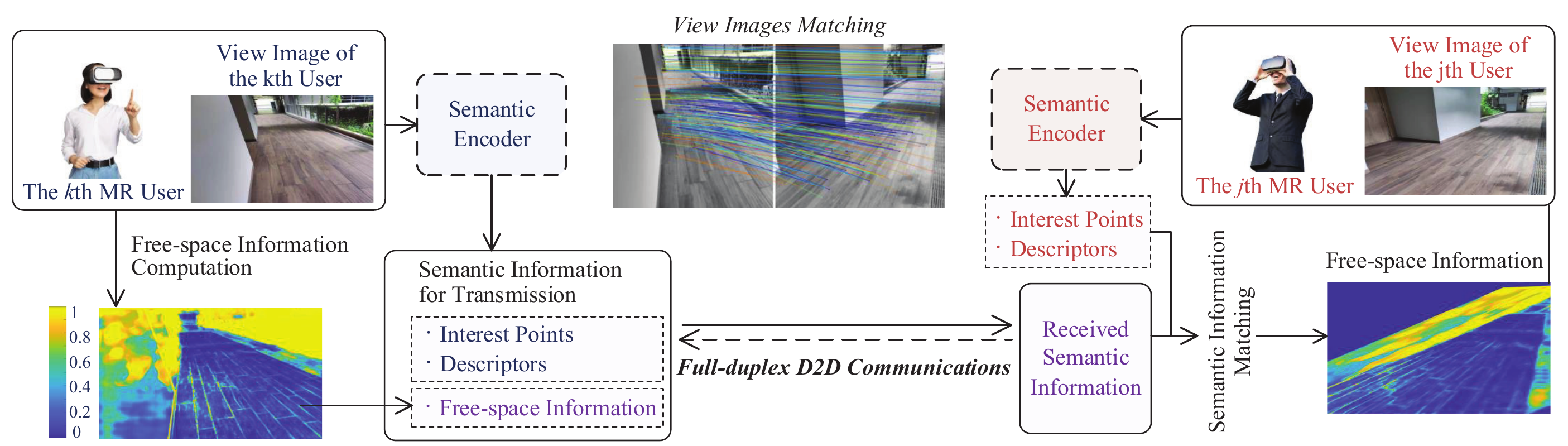}%
\caption{The semantic-aware free-space information sharing mechanism proposed in this paper. Full-duplex wireless communication is performed between MR HMDs among users. The free-space information is computed by the state-of-art anomaly detection approach named JSR-Net~\cite{vojir2021road}}
\label{model}
\end{figure*}
When users use an MR technique to access virtual services, e.g., Metaverse, the MR HMDs need to perform various computing tasks. As shown in Fig.~\ref{ready}, HMD devices should compute positions of virtual objects superimposed on the real world, and the planning of routes that users can walk safely in real time. Since even the perspectives of users in the same scene may differ, each user needs to perform the above tasks independently. For example, the $k^{\rm th}$ user generates a virtual flower on the ground, and the $j^{\rm th}$ user, to synchronize this operation, needs to compute where the flower is on the ground according to the $j^{\rm th}$ user's own perspective. A possible solution is to perform user view sharing as shown in Fig.~\ref{fwmodel}. The shared users can obtain the sharer's computation results by view matching without repeating the computation independently. Still with the above example, the $k^{\rm th}$ user can analyze and than share his own view information to other users with similar views, and other users can determine the location of the flower through the view matching. However, the sharing of original view images among MR HMDs consumes a lot of wireless transmission resources, i.e., transmit power and bandwidth. The reason is that the HMD resolution is high, resulting in a large amount of data for view images.

To solve this problem and achieve efficient and effective information sharing, we apply the semantic communication technique in the D2D full-duplex communication system. A representative task could be {\textit{free-space information sharing}}. As shown in Fig.~\ref{model}, specifically, a user who identifies an area in their view image that is safe to walk in can share the detected free-space information with other users. Other users can then directly perform the matching to obtain safe walking area in their own view images. Let ${\bf{x}}_k$ denote the original view image of the $k^{\rm th}$ user. Here the semantic information includes the detected free-space (${\bf{i}}_k$), the interest points that reflect important structures in the visual field diagram (${{\bf{p}}_k}$), and the descriptors corresponding to interest points (${{\bf{d}}_k}$). Let $\mathcal{S}\left\{ {\cdot} \right\}$ denote the semantic encoder, we have
\begin{equation}
\mathcal{S}\left\{ {{{\bf{x}}_k}} \right\} = \left\{ {{{\bf{p}}_k},{{\bf{d}}_k},{{\bf{i}}_k}} \right\}.
\end{equation}
The total data size of the semantic information, i.e., $\left\{ {{p_k},{d_k},{i_k}} \right\}$, will be much smaller than that of the original view image, i.e., $\left\{ {{x_k}} \right\}$. The detailed design of the semantic encoder is presented in Section~\ref{S4A}.

\subsection{Full-Duplex Wireless Communications}\label{wirelesssec}
Although semantic communication technologies can reduce the amount of data transmitted without affecting task completion, bandwidth resources are still strained when the number of users is large. Because of limited bandwidth resources in the HMD device network, we consider that users are communicating with each other in the full-duplex mode.

A critical issue in full-duplex communication is self-interference cancellation. To investigate the effect of the residual self-interference, i.e., ${{I}_k}$, on system performance, we consider that ${{I}_k}$ is subject to the Gaussian distribution~\cite{kong2022neural}. This can be regarded as the worst-case assumption~\footnote{The exact residual self-interference modeling could be different according to the interference cancellation technique, which is left for our future work.} about the interference~\cite{han2020spectrum}. Specifically, the residual self-interference at $k^{\rm th}$ user and $j^{\rm th}$ user can be modeled as zero-mean complex Gaussian random variables with variance ${\upsilon _k}{P_k} \sigma_S^2$ and ${\upsilon _j}{P_k} \sigma_S^2$, respectively~\cite{han2020spectrum}, where ${P_k}$ is the transmit power of the $k^{\rm th}$ user and $\sigma_S^2$ denotes the variance due to unit power. The parameters ${\upsilon _k}$ and ${\upsilon _j}$ are the constants that reflect the self-interference cancellation abilities of $k^{\rm th}$ user and $j^{\rm th}$ user, respectively~\cite{han2020spectrum}. 

We consider that the $j^{\rm th}$ user is transmitting the semantic information to the $k^{\rm th}$ user. Let $s_j$ denote the transmitted symbol in $\mathcal{S}\left\{ {{{\bf{x}}_j}} \right\}$. The received signal at the $k^{\rm th}$ user can be expressed as
\begin{equation}
{{y}_k} = \sqrt {{{P}_j}D_{jk}^{ - {\beta _k}}} {h_{jk}}{{s}_j} + \sum\limits_{i = 1}^N {\sqrt {{P_I}} } {g_{k}}{w_i} + {{I}_k} + {{n}_k},
\end{equation}
where $ {n_k} \sim \mathcal{C}\mathcal{N}\left( {0,\sigma _N^2} \right) $, $ {I_k} \sim \mathcal{C}\mathcal{N}\left( 0 ,{\upsilon_k}{P_k} \sigma_S^2 \right) $, $h_{jk}$ is the small-scale fading channel, ${P_j}$ is the transmit power of the $j^{\rm th}$ user, $D_{jk}$ is the distance between the $k^{\rm th}$ user and $j^{\rm th}$ user, $\beta_{k}$ is the corresponding path loss exponents, $N$ is the number of interfere paths that are assumed to be present at the $k^{\rm th}$ user-$j^{\rm th}$ user full-duplex communications pair, each of the interfering signals has an average transmit power $P_{I}$, {{${w_{i}}$}} is the $i^{\rm th}$ interfering symbol, and $g_k$ is the small-scale fading of the interfering signal.

The frequency band used for D2D communications among users is lower than 5 GHz. We consider that the interference signals follow the Rayleigh distribution~\cite{durgin2002new}, i.e., $ {{{\left| {g_{k}} \right|}^2}} \sim {\rm Rayleigh} \left( {\eta _{k}}\right)$, which is practical and widely considered in many works~\cite{gao1998theoretical,kang2022personalized}. Note that the large scale fading of the interference signal is considered in the mean value of ${g_{k}}$. We then use the $\alpha-\mu$ distribution to model the small-scale fading, which is a general fading model that includes several important other distributions, such as the Weibull, One-Sided Gaussian, Rayleigh, and Nakagami. The probability density function (PDF) and the cumulative distribution function (CDF) expressions of a squared $\alpha - \mu$ random variable ${{\left| {{h_{jk}}} \right|}^2}$ are given by~\cite{yacoub2007alpha}:
\begin{equation}\label{PDFHJK}
{f_{{\left| {{h_{jk}}} \right|}^2}}\left( x  \right) = \frac{{\alpha {x ^{\frac{{\alpha \mu }}{2} - 1}}}}{{2{\beta ^{\frac{{\alpha \mu }}{2} }}\Gamma \left( \mu  \right)}}\exp \left( { - {{\left( {\frac{x}{\beta }} \right)}^{\frac{\alpha }{2}}}} \right),
\end{equation}
and
\begin{equation}
{F_{{\left| {{h_{jk}}} \right|}^2}}\left( x  \right) = \frac{{\gamma\left( {\mu ,{x ^{\frac{\alpha }{2}}}{\beta ^{ - \frac{\alpha }{2}}}} \right)}}{{\Gamma \left( \mu  \right)}},
\end{equation}
respectively, where {$\Gamma\left(\cdot \right) $} is the gamma function \cite[eq. (8.310.1)]{gradshteyn2007}, {$ \beta  = \frac{{\bar \Upsilon \Gamma \left( \mu  \right)}}{{\Gamma \left( {\mu  + \frac{2}{\alpha }} \right)}} $}, and $\gamma\left(\cdot \right) $ is the incomplete gamma function \cite[eq. (8.35)]{gradshteyn2007}.

The received signal-to-interference-plus-noise Ratio (SINR) of the $k^{\rm th}$ user can be expressed as
\begin{equation}\label{SINR}
{\gamma _k} = \frac{{{P_j}D_{jk}^{ - {\beta _k}}{{\left| {{h_{jk}}} \right|}^2}}}{{{P_I}\sum\limits_{i = 1}^N {{{\left| {g_{k}} \right|}^2}}  + {\upsilon_k}{P_k}\sigma _S^2 + \sigma _N^2}}.
\end{equation}
Simply, the SINR received by the $j^{\rm th}$ user can be obtained by interchanging $j$ and $k$ in \eqref{SINR}. Note that the value of SINR affects the BEP and achievable rate in the wireless semantic information transmission. In full-duplex communications, the higher the transmit power of one user, the larger the value of SINR of the received signal of the other user. Considering that the total energy of the user's MR HMD is limited, incentives can be used to encourage users to share information, which is discussed in Section~\ref{ser2}.

\section{Semantic Matching and Contract Modelling}\label{S4}
In this section, we propose the semantic encoder design scheme according to SuperPoint~\cite{detone2018superpoint} and the semantic matching scheme according to SuperGlue~\cite{sarlin2020superglue}, using the received semantic information to perform view images matching.
\subsection{Self-Supervised Semantic Encoding}\label{S4A}
% Super Point
The definition of semantic information is typically task-relevant. For source messages in image modality, semantic information can be in the form of knowledge graphs~\cite{ji2021survey}, semantic segmentation results~\cite{ng2022stochastic}, or features~\cite{kang2022task}. In our proposed D2D free-space information sharing framework, the semantic information extracted from one user's view image needs to be used to help a second user determine the spatial correspondence between two view images.

\begin{figure*}[t]
\centering
\includegraphics[width=0.85\textwidth]{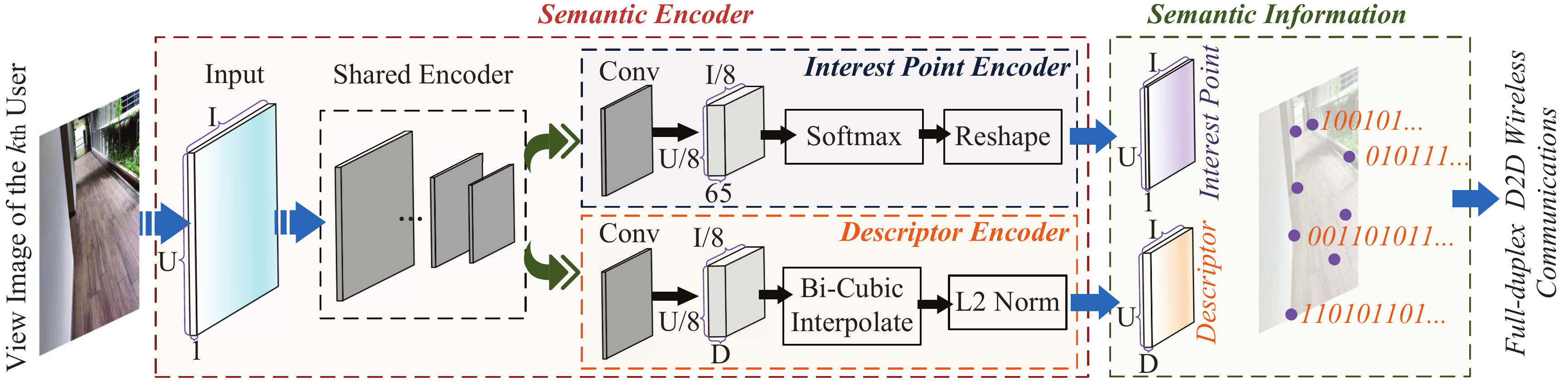}%
\caption{The structure of the semantic encoder. The input is the user's view image, and the output is the detected interest points and descriptors.}
\label{SemanticEncoder}
\end{figure*}
As we discussed in Section~\ref{slam}, convolutional neural network (CNN) has been widely used in interest point detection and description. Here, we utilize the self-supervised SuperPoint~\cite{detone2018superpoint} architecture as the semantic encoder to produce interest point detection accompanied by fixed length descriptors in a single forward pass. Specifically, as shown in Fig.~\ref{SemanticEncoder}, the full-sized view image of the $k^{\rm th}$ user is sent to a shared CNN to reduce the dimensionality of the input view image. The output is then fed to two separate encoders to obtain the interest points and descriptors, respectively. We next present the introduction to each component and the design of the loss function.

\begin{itemize}
\item {\textit{Shared CNN Encoder:}} To reduce the dimensionality of the input view image, the shared encoder uses the VGG-style~\cite{simonyan2014very} architecture. The encoder consists of convolutional layers, spatial down-sampling via pooling and non-linear activation functions~\cite{detone2018superpoint}. The encoder maps an input image to an intermediate tensor with smaller spatial dimension and greater channel depth.
\item {\textit{Interest Point Encoder:}} The interest point encoder in the semantic encoder can be designed with an explicit decoder to reduce the computation of the model~\cite{odena2016deconvolution}. After a channel-wise softmax layer, the dustbin dimension is removed and the reshape is performed.
\item {\textit{Descriptor Encoder:}} A model similar to UCN~\cite{choy2016universal} is used in the descriptor encoder to first output a semi-dense grid of descriptors. The decoder then performs bicubic interpolation of the descriptor. The activation is $L2$-normalized to be unit length, as shown in Fig.~\ref{SemanticEncoder}.
\item {\textit{Loss Function Design:}} The loss function of the semantic encoder can be written as the sum of the loss functions of the two encoders, i.e., interest point and descriptor encoders. According to the SuperPoint~\cite{detone2018superpoint}, pairs of synthetically warped images can be used to train the network.
\end{itemize}

After the network is trained, the required interest points and descriptors can be quickly obtained by inputting the user's view image. It has been shown that SuperPoint-based network architecture can support real-time interest points and descriptors extraction~\cite{detone2018superpoint}.

The extracted interest points and descriptors are received by other users via D2D full-duplex wireless transmission and then used. The performance metrics in wireless transmission are discussed in Section~\ref{S5}. We first discuss the semantic information matching problem after successful reception in the following.

\subsection{Semantic Information Matching}\label{smatch}
\begin{figure*}[t]
\centering
\includegraphics[width=0.8\textwidth]{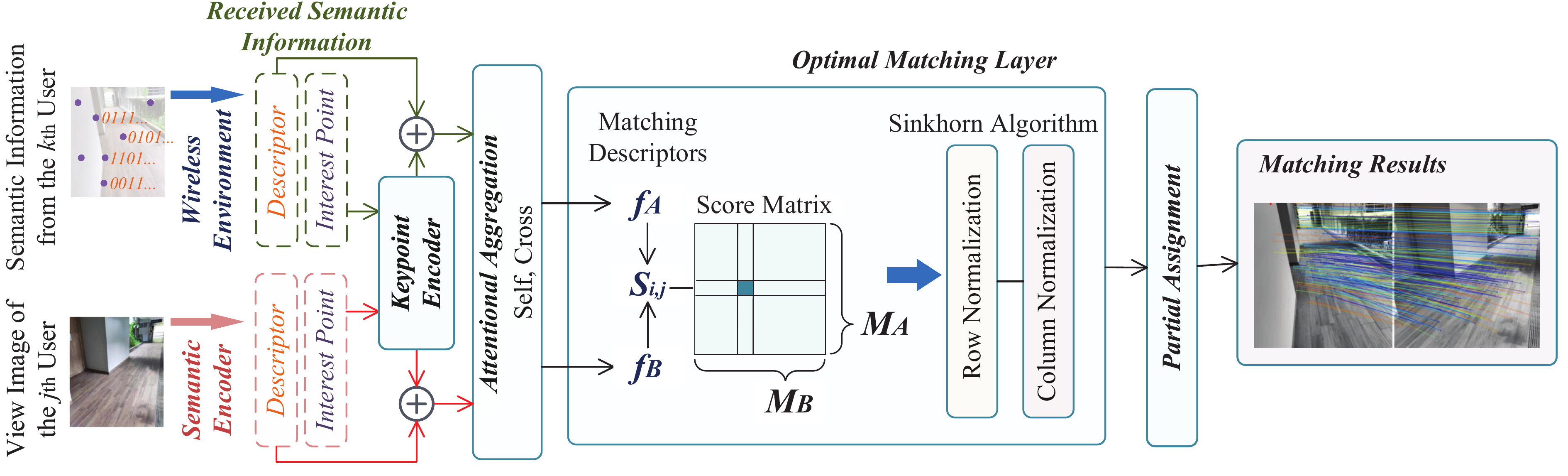}%
\caption{The structure of the semantic matching. The input is the received semantic information and the $j^{\rm th}$ user's semantic encoding results, and the output is the matching results.}
\label{SemanticMatch}
\end{figure*}
In the full-duplex system under our consideration, the semantic encoders discussed in the previous section should be deployed in each MR HMD, as shown in Fig.~\ref{model}. After other users receive the interest points and descriptors, semantic information extraction should also be performed on the user's own view image. Then, the semantic information matching is performed among the extracted interest points and descriptors. Here, we use SuperGlue~\cite{sarlin2020superglue} as a semantic matching network architecture. Note that a user may share its semantic information to multiple users in the form of D2D full-duplex wireless communication, each user is independent when matching semantic information. Therefore, in the following discussion we focus on the matching between the two view images.

Here, the semantic information of one image, i.e., interest point and descriptors, obtained by the semantic encoder can be regarded as the local features. We consider that images $A$ and $B$ have $M_A$ and $M_B$ local features, respectively.
According to the SuperGlue~\cite{sarlin2020superglue}, for the integration into downstream tasks and better interpretability, each possible correspondence should have a confidence value. We consequently define a partial soft assignment matrix ${\bf{P}}\in {\left[ {0,1} \right]^{M_A \times M_B}}$ as ${\mathbf{P}}{{\mathbf{1}}_{M_B}} \leqslant {{\mathbf{1}}_{M_A}}$, and ${{\mathbf{P}}^{\text{T}}}{{\mathbf{1}}_{M_A}} \leqslant {{\mathbf{1}}_{M_B}}$. The goal is to design a neural network that predicts the assignment ${\bf{P}}$ from two sets of local features. In the following, we introduce the main components and the design of the loss function.
\begin{itemize}
\item {\textit{Attentional Graph Neural Network:}} To enhance the uniqueness of points of interest and descriptors, we can consider the relationship between a certain pair of points of interest and descriptors and other pairs, such as the ones that are salient, self-similar, statistically co-occurring, or adjacent~\cite{trzcinski2018scone}. Also, notice that when humans are asked to match a given pair blurred key points, the humans look back and forth on both images~\cite{chun2000contextual}. This implies that an iterative process can focus the attention on a specific correct location.
Thus, the first major block of SuperGlue is designed as an Attentional Graph Neural Network (AGNN)~\cite{sarlin2020superglue}.
For given initial local features, the AGNN computes matching descriptors by allowing features interact with other features. We embed the keypoint position into a high dimensional vector with a Multilayer Perception (MLP), which enables the graph network to later reason about both appearance and position jointly, especially when combined with attention~\cite{gehring2017convolutional}.
\item {\textit{Optimal Matching Layer:}} The optimal matching layer is the second major block of SuperGlue. It is used to produce a partial assignment matrix. As in the standard graph matching formulation, the assignment $\mathbf P$ can be obtained by computing a score matrix $\mathbf S \in \mathbb{R}^{M_A\times M_B}$ for all possible matches and maximizing the total score $\sum_{i,j}\mathbf S_{i,j} \mathbf P_{i, j}$. This is equivalent to solving a linear assignment problem. The optimal matching layer creates an ${M_A\times M_B}$ score matrix, augments it with dustbins, and then finds the optimal partial assignment by using the Sinkhorn algorithm~\cite{peyre2019computational}. After $T$ iterations, we drop the dustbins and recover $\mathbf{P} = \bar{\mathbf{P}}_{1:M_A,1:M_B}$.
\item {\textit{Loss Function Design:}} Both the AGNN and the optimal matching layer are differentiable, which allows back propagation from matches to visual descriptors. Note that SuperGlue is trained in a supervised manner from ground truth matches~\cite{sarlin2020superglue}. Given these labels, we minimize the negative log-likelihood of the assignment $\bar{\mathbf P}$. This supervision aims at simultaneously maximizing the precision and the recall of the matching. After training, the model can be used in the user's MR HMDs to match semantic information, after receiving semantic features from the transmitter and extracting its own interest points and descriptors from the view image.
\end{itemize}

\subsection{Contract Modelling}
We use a contract as the incentive to encourage users to share the semantic information with each other. The payment plan and the utilities of the SIP and SIR are formulated respectively.

\subsubsection{Payment Plan}
To promote MR users' participation in the semantic information sharing, an appropriate payment plan is required that allows both SIP and SIR benefit from the cooperation. To this end, we propose a contract theoretic payment plan, in which the SIP receives the compensation from the SIR according to quality of the shared semantic information (QoS). Thus, the resulting revenue function of the InP can be expressed as
\begin{equation}
{I_{{\rm{SIP}}}} = c_q Q_s + c_f,
\end{equation}
where $c_f$ represents the payment made by the SIR to the SIP for the computation resource consumption in semantic encoding, the parameter $c_q$ denotes the fee charged for a unit QoS value, and {$Q_s$} is the quantified value of the QoS of the SIR. 

While the proposed semantic information extraction and matching schemes in Sections~\ref{S4A} and \ref{smatch} demonstrate efficiency and feasibility, the semantic matching outcomes are subject to the wireless transmission process. The achievable rate impacts the latency of the information sharing, while the BEP affects the accuracy of the received semantic information. Hence, we define the QoS as
\begin{equation}
Q_s (P_k)  = M_k \mathcal{T}\left(R_{k}\right) \mathcal{T}\left(1 - E_{k}\right),
\end{equation}
where $M_k \in [0,1]$ indicate the impact of the semantic extraction and matching algorithm on QoS\footnote{Because $M_k$ is only decided by the chosen semantic encoding and matching algorithms, we set $M_k = 1$ without loss of generality.}, $R_{k}$ is the achievable rate, $E_{k}$ is the BEP, and the function {\small $ {\cal T}\left(  \cdot  \right) $} is used to eliminate the effect of the magnitudes~\cite{nishio2013service}, which is defined as
\begin{equation}\label{calT}
{\cal T}\left( t \right) = \frac{{t - {t_{\min }}}}{{{t_{\max }} - {t_{\min }}}},
\end{equation}
where $ {{t_{\min }}} $ is the minimal threshold, $ {{t_{\max }}} $ is the maximal value that SIP can provide.

\subsubsection{Utility of The Semantic Information Provider}
The utility of the SIP can be obtained as
\begin{equation}\label{SIP}
{U_{\rm SIP}} = {I_{{\rm{SIP}}}} - c_p P_k = c_q Q_s - c_p P_k + c_f,
\end{equation}
where $P_k$ is the transmit power of the SIP and $c_p$ represents the unit cost of $P_k$.

\subsubsection{Utility of The Semantic Information Receiver}
As we discussed in Section~\ref{yuanli}, SIR can match spatial locations with the SIP using the received semantic information. Therefore, the higher the $Q_s$, the more accurate the results achieved by SIR through matching, and the more computation resources can be saved. We use the parameter $c_s$ to measure the utility gain per unit $Q_s$ boost to SIR. The utility of the SIR can be then expressed as
\begin{equation}\label{SIR}
{U_{\rm SIR}} = c_s Q_s - {I_{{\rm{SIP}}}} = \left( c_s-c_q\right)  Q_s - c_f.
\end{equation}

\subsubsection{Contract Modelling}\label{contractm}
The contract provided by the SIR includes two items, i.e., {\small $ \left\{ {{c_q},{c_f}} \right\} $}. To design the optimal contract, we formulate the SIR's utility maximization problem while providing the SIP with the necessary incentives to agree on the contract. The optimization problem can be expressed as~\cite{zhang2017multi}
\begin{equation}\label{qu}
\begin{array}{*{20}{c}}
{\mathop {\max }\limits_{{{c_q},{c_f},{P_k}}} }&{{U_{\rm SIR}} \left( {{c_q},{c_f},{P_k}} \right)}\\
{\rm{s.t.}}&{\left\{ \begin{array}{l}
{{P_k}^*} \in \arg \mathop {\max }\limits_{P_k} {U_{{\rm{SIP}}}}\left( {{P_k},{c_q},{c_f}} \right),\\
{U_{{\rm{SIP}}}}\left( {{{P_k}^*},{c_q},{c_f}} \right) \ge U_{\rm th}^{\rm SIP},
\end{array} \right.}
\end{array}
\end{equation}
where the first constraint is the IC constraint~\cite{zhang2017contract}, i.e., {\small ${P_k}^*$} is set to maximize its own utility. The second is the IR constraint~\cite{zhang2017contract} with a utility threshold {\small $U_{\rm th}^{\rm SIP}$}.

We first consider the IC constraint. The closed-form expression of QoS can be derived in Section~\ref{S5}. For a given contract $ \left\{ {c_q},{c_f} \right\} $, we can solve the IC constraint to obtain the optimal power and corresponding $Q_s^*$. Accordingly, we substitute the IR constraint with the optimal power and simplify the SIR maximization problem. Therefore, the optimal contract design, {\small $ \left\{ {{c_q}^*,{c_f}^*} \right\} $}, is modeled as a complex decision-making problem. Whereas conventional mathematical techniques may encounter tractability challenges in addressing a given problem, DRL-based methods might offer a viable alternative.

However, the optimal contract design for a wireless communication system is impacted not only by user-specific factors such as price per unit of power, but also by the intricate wireless milieu, including co-channel interference signal power, the count of interference paths, and different parameters associated with small-scale fading as discussed in Section~\ref{wirelesssec}. As a result, the DRL-based algorithms may struggle to converge to the optimal solution or achieve suboptimal performance due to the high dimensionality of the state. To address this challenge, the choice of DRL algorithms can be crucial to improve the effectiveness. Proximal Policy Optimization (PPO) and Soft Actor-Critic (SAC) are two popular DRL algorithms that have shown promising results in high-dimensional state problems. PPO is an on-policy algorithm that optimizes the policy by iteratively updating the policy using a clipped surrogate objective. It prevents large policy updates and promotes smooth policy changes. On the other hand, SAC is an off-policy algorithm that learns a stochastic policy to maximize both the expected cumulative reward and the entropy of the policy. SAC is known for its ability to handle continuous action spaces and can achieve state-of-the-art performance in a variety of tasks.

Therefore, we solve the optimal contract design problem, i.e., \eqref{qu}, using two DRL algorithms, i.e., PPO and SAC. Moreover, to tackle the challenge of addressing high-dimensional state spaces using DRL algorithms, we propose a novel AI-generated contract method based on diffusion models in Section~\ref{ser2}. We then evaluate and compare the performance of our proposed method with that of the DRL schemes in Section~\ref{difftest}.

\section{AI-Generated Contract And Performance Analysis}\label{S5}
In this section, we proposed an AI-generated contract method. Then, we derive the PDF expression of the SINR in the D2D full-duplex communication system. The closed-form achievable rate and BEP are also derived.

\subsection{AI-Generated Contract}\label{ser2}{
\subsubsection{Diffusion Model}
Diffusion-based generative models pose the data-generating process as an iterative denoising procedure. This denoising is the opposite of a forward diffusion process, which gradually corrupts the data's structure by introducing noise. Specifically, diffusion models are latent variable models of the form
\begin{equation}\label{agfdsea}
{p_\theta }\left( {{{\mathbf{x}}_0}} \right): = \int {{p_\theta }} \left( {{{\mathbf{x}}_{0:T}}} \right)d{{\mathbf{x}}_{1:T}},
\end{equation}
where $ {{\mathbf{x}}_1}, \ldots ,{{\mathbf{x}}_T} $ are latents of the same dimensionality as the data ${{\mathbf{x}}_0 \sim p\left({\mathbf{x}}_0\right) }$. The joint distribution in~\eqref{agfdsea}, i.e., ${{p_\theta }} \left( {{{\mathbf{x}}_{0:T}}} \right)$, is called the {\textit{reverse diffusion}}, which is defined as a Markov chain with learned Gaussian transitions starting at $ p\left( {{{\mathbf{x}}_T}} \right) = \mathcal{N}\left( {{{\mathbf{x}}_T};{\mathbf{0}},{\mathbf{I}}} \right) $ as
\begin{equation}
{p_\theta }\left( {{{\mathbf{x}}_{0:T}}} \right): = p\left( {{{\mathbf{x}}_T}} \right)\prod\limits_{t = 1}^T {{p_\theta }} \left( {{{\mathbf{x}}_{t - 1}}\mid {{\mathbf{x}}_t}} \right),
\end{equation}
where 
\begin{equation}
{p_\theta }\left( {{{\mathbf{x}}_{t - 1}}\mid {{\mathbf{x}}_t}} \right): = \mathcal{N}\left( {{{\mathbf{x}}_{t - 1}};{{\mathbf{\mu }}_\theta }\left( {{{\mathbf{x}}_t},t} \right),{{\mathbf{\Sigma }}_\theta }\left( {{{\mathbf{x}}_t},t} \right)} \right).
\end{equation}
A {\textit{forward diffusion}} chain gradually adds noise to the data as a Markov chain according to a variance schedule $ {\beta _1}, \ldots ,{\beta _T} $ as
\begin{equation}
q\left( {{{\mathbf{x}}_{1:T}}\mid {{\mathbf{x}}_0}} \right): = \prod\limits_{t = 1}^T q \left( {{{\mathbf{x}}_t}\mid {{\mathbf{x}}_{t - 1}}} \right),
\end{equation}
where 
\begin{equation}
q\left( {{{\mathbf{x}}_t}\mid {{\mathbf{x}}_{t - 1}}} \right): = \mathcal{N}\left( {{{\mathbf{x}}_t};\sqrt {1 - {\beta _t}} {{\mathbf{x}}_{t - 1}},{\beta _t}{\mathbf{I}}} \right).
\end{equation}
After training, sampling from the diffusion model consists of sampling $\boldsymbol{x} \sim p\left(x_T\right)$ and running the reverse diffusion chain to go from $t = T$ to $t = 0$. Diffusion models can be straightforwardly extended to conditional models by conditioning $ {p_\theta }\left( {{{\mathbf{x}}_{t - 1}}\mid {{\mathbf{x}}_t},c} \right) $.

\subsubsection{AI-Generated Contract}
\begin{figure*}[t]
\centering
\includegraphics[width=0.9\textwidth]{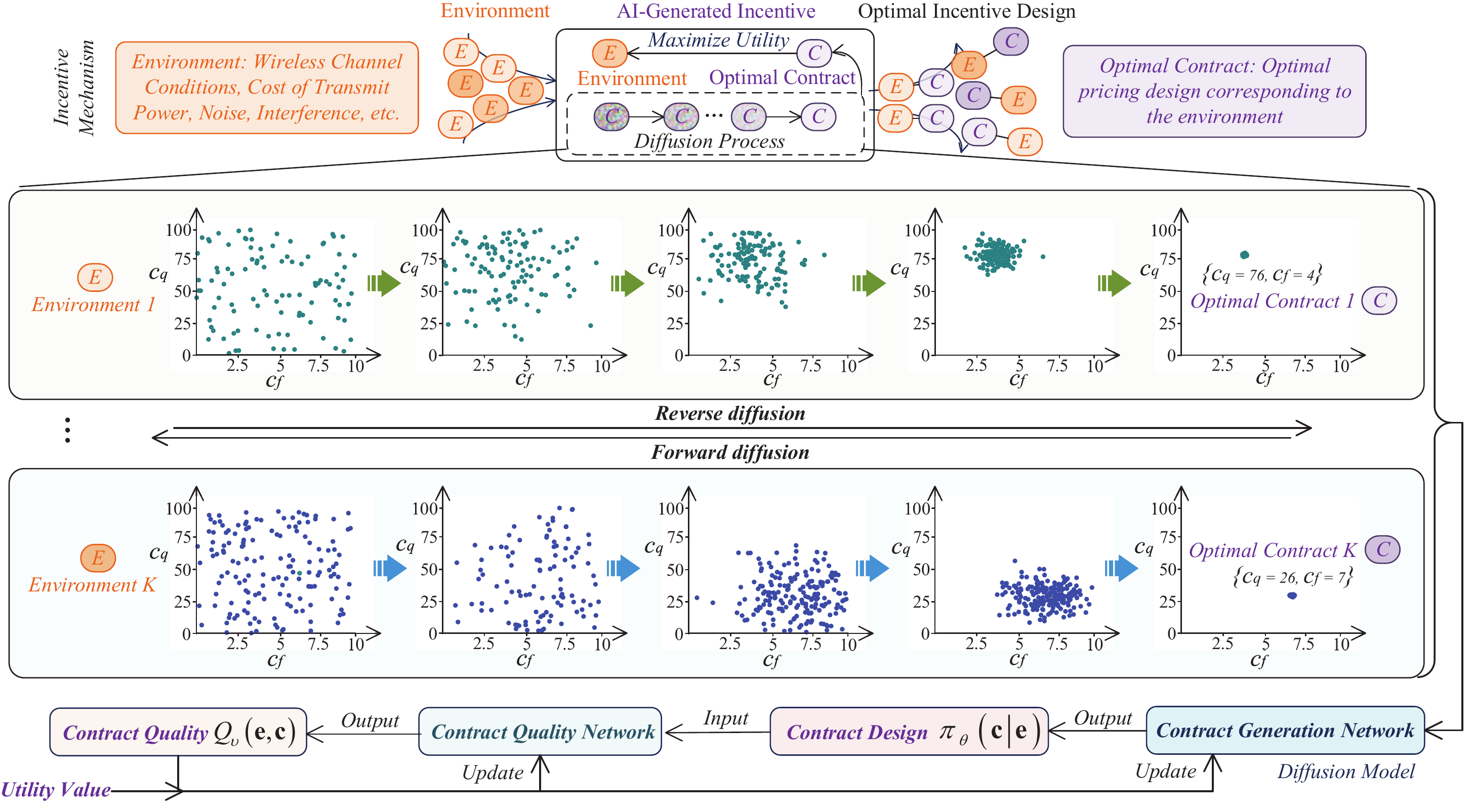}%
\caption{Diffusion Model for incentive design. Diffusion Model for incentive design. For any given environment, the contract is first randomly generated as the Gaussian noise. After the multi-step denoising through the diffusion model, the output is the contract design that maximizes the optimization objective.}
\label{incentive}
\end{figure*}
We explain how to apply a conditional diffusion model to generate the contract design, i.e., $\left\{c_q, c_f\}\right.$ in \eqref{qu}. In contrast to the conventional utilization of back propagation algorithms in neural networks or DRL techniques that directly optimize model parameters, diffusion models aim to enhance contract design by iterative denoising the initial distribution, leading to generating a utility function of greater efficacy.

As we discussed in Section~\ref{contractm}, the optimal contract is related to multiple variables that are unknown to the SIR. We use the vector ${\bm{e}}$ to denote the environment, i.e., the set of all variables, that affects the optimal contract design as
\begin{equation}\label{env}
{\bm{e}} = \left\{ {{c_s},{c_p},{D_{jk}},{\beta _k},\alpha ,\mu ,\sigma _N^2,\sigma _S^2,{\upsilon _k},{P_I},{\eta _k},N,{P_j}} \right\}.
\end{equation}
We denote the contract to be designed by the SIR in the given environment as ${\bm{c}} = {\left\{c_q,c_f\right\}}$. We call the diffusion model network that maps environment states to contract designs as the {\textit{contract design policy}}, i.e., ${\pi _\theta }\left( {\left. {\bm{c}} \right|{\bm{e}}} \right)$ with parameters $\theta$. The goal of ${\pi _\theta }\left( {\left. {\bm{c}} \right|{\bm{e}}} \right)$ is to output deterministic contract design that maximizes the expected cumulative reward over a sequence of time steps. We represent the {\textit{contract design policy}} via the reverse process of a conditional diffusion model as
\begin{equation}
{\pi _\theta }\!\left( {\left. {\bm{c}} \right|{\bm{e}}} \right) = {p_\theta }\!\left( {\left. {{{\bm{c}}^{0:N}}} \right|{\bm{e}}} \right) = \mathcal{N}\!\left( {{{\bm{c}}^N};{\bm{0}},{\bm{I}}} \right)\prod\limits_{i = 1}^N {{p_\theta }\!\left( {\left. {{{\bm{c}}^{i - 1}}} \right|{{\bm{c}}^i},{\bm{e}}} \right)}.
\end{equation}
As shown in Fig.~\ref{incentive}, the end sample of the reverse chain is the final chosen contract design. Here, ${{p_\theta }\left( {\left. {{{\bm{c}}^{i - 1}}} \right|{{\bm{c}}^i},{\bm{e}}} \right)}$ can be modeled as a Gaussian distribution $ \mathcal{N}\left( {{{\mathbf{c}}^{i - 1}};{{\mathbf{\mu }}_\theta }\left( {{{\mathbf{c}}^i},{\mathbf{e}},i} \right),{{\mathbf{\Sigma }}_\theta }\left( {{{\mathbf{c}}^i},{\mathbf{e}},i} \right)} \right) $. According to~\cite{ho2020denoising}, ${{p_\theta }\left( {\left. {{{\bm{c}}^{i - 1}}} \right|{{\bm{c}}^i},{\bm{e}}} \right)}$ can be modeled as a noise prediction model with the covariance matrix fixed as
\begin{equation}
\boldsymbol{\Sigma}_\theta\left(\boldsymbol{c}^i, \boldsymbol{e}, i\right)=\beta_i \boldsymbol{I},
\end{equation}
and mean constructed as
\begin{equation}
{{\mathbf{\mu }}_\theta }\left( {{{\mathbf{c}}^i},{\mathbf{e}},i} \right) = \frac{1}{{\sqrt {{\alpha _i}} }}\left( {{{\mathbf{c}}^i} - \frac{{{\beta _i}}}{{\sqrt {1 - {{\bar \alpha }_i}} }}{{\bm{\varepsilon}} _\theta }\left( {{{\mathbf{c}}^i},{\mathbf{e}},i} \right)} \right),
\end{equation}
where $\alpha_i = 1 - \beta_i$ and ${\bar \alpha}_i = \prod\limits_{s = 1}^i a_s$. 
We first sample $ {{\bm{c}}^N}\sim\mathcal{N}({\mathbf{0}},{\mathbf{I}}) $ and then from the reverse diffusion chain parameterized by $ \theta  $ as
\begin{equation}\label{denoise}
{{\mathbf{c}}^{i - 1}}\mid {{\mathbf{c}}^i} = \frac{{{{\mathbf{c}}^i}}}{{\sqrt {{\alpha _i}} }} - \frac{{{\beta _i}}}{{\sqrt {{\alpha _i}\left( {1 - {{\bar \alpha }_i}} \right)} }}{{\bm{\varepsilon}} _\theta }\left( {{{\mathbf{c}}^i},{\mathbf{e}},i} \right) + \sqrt {{\beta _i}} {\bm{\varepsilon}}.
\end{equation}
Therefore, the task of training the \textit{contract design policy} $\pi_\theta$ in complex and high-dimensional environments ${\bm e}$ is effectively transferred to training a \textit{contract generation network} ${{\bm{\varepsilon}}_\theta }$. Following DDPM~\cite{ho2020denoising}, when $i = 1$, ${\bm{\varepsilon}}$ is set as $0$ to improve the sampling quality. Furthermore, to train the ${\bm{\varepsilon}}_\theta$, motivated by the Q-function in DRL, we define the {\textit{contract quality network}} $Q_\upsilon$ that maps an environment-contract pair, i.e., $\{{\bm e}, {\bm c}\}$, to a value representing the expected cumulative reward if an agent takes one contract design policy from the current state and follows the policy thereafter. Thus, the optimal {\textit{contract design policy}} is the policy that maximizes the expected cumulative utility of the SIR, which can be obtained by
\begin{equation}
\pi  = \mathop {\arg \min }\limits_{{\pi _\theta }} \mathcal{L}(\theta ) =  - {\mathbb{E}_{{{\mathbf{c}}^0}\sim{\pi _\theta }}}\left[ {{Q_\upsilon }\left( {{\mathbf{e}},{{\mathbf{c}}^0}} \right)} \right].
\end{equation}
%{\mathbf{e}}\sim\mathcal{D}
The \textit{contract quality network} is learned in a conventional way, minimizing the Bellman operator with the double Q-learning technique~\cite{hasselt2010double}. Two networks, namely $ {{Q_{{\upsilon_1}}}} $ and $ {{Q_{{\upsilon_2}}}} $, along with target networks, namely $ {{Q_{\upsilon_1^\prime }}} $, $ {{Q_{\upsilon_2^\prime }}} $ and $ {{\pi _{{\theta ^\prime }}}} $, are constructed. We then optimize $\upsilon_i$ for $i = \left\{1,2 \right\} $ by minimizing the objective
\begin{equation}
{\mathbb{E}_{{\mathbf{c}}_{t + 1}^0\sim{\pi _{{\theta ^\prime }}}}}\!\!\!\left[ {{{\left\| {\!\left(\! {r\!\left( {{\mathbf{e}},{{\mathbf{c}}_t}} \right) \!+\! \gamma {{\mathop {\min }\limits_{i = 1,2} }}{Q_{\upsilon_i^\prime }}\!\!\left( {{\mathbf{e}},{\mathbf{c}}_{t + 1}^0} \right)} \!\right) \!\!- \! {Q_{{\upsilon_i}}}\!\!\left( {{\mathbf{e}},{{\mathbf{c}}_t}} \right)} \right\|}^2}} \right].
\end{equation}
Specifically, our AI-generated contract algorithm involves generating a contract design using the denoising technique, followed by adding exploration noise to the contract design and executing it to gain the exploring experience. The algorithm for AI-generated contract is shown in {\bf{Algorithm~\ref{Algorithm}}}. By utilizing {\bf{Algorithm~\ref{Algorithm}}}, we can obtain the optimal contract based on the wireless system environment vector. The performance of the proposed AI-generated contract algorithm is compared with conventional DRL algorithms in Section~\ref{S6}.

\begin{rem}
Our proposed diffusion model-based AI-generated contract algorithm learns experience from the environment in an exploratory manner and achieves the optimal solution using the diffusion model. By gradually adjusting the output, the diffusion model is more robust and effective than the neural network model that directly outputs the solution. We believe that our proposed innovative algorithm can be easily extended to various problems in wireless communication networks, such as resource allocation, game theoretic incentive mechanism design, and reinforcement learning and federal learning algorithm optimization.
\end{rem}

In the following, we further analyze the performance of the full-duplex wireless system and provide more insights.

\begin{algorithm}[t]
{\small \caption{The Algorithm for AI-Generated Contract}
\label{Algorithm}
\hspace*{0.02in} {\bf{\textit{Training Phase:}}}
\begin{algorithmic}[1]
\State Input hyper-parameters: diffusion step $N$, batch size $N_b$, discount factor $\gamma$, soft target update parameter $\tau$, exploration noise $\epsilon$
\vspace{0.1cm}
\State {\textit{\#\# \quad Initialize NN}}
\State Initialize replay buffer $R$, contract generation network ${\bm{\varepsilon}}_\theta$ with weights $\theta$, contract quality network $Q_\upsilon$ with weights $\upsilon$, target contract generation network ${\bm{\varepsilon}'}_{\theta'}$ with weights ${\theta'}$, target contract quality network ${Q'}_{\upsilon'}$ with weights $\upsilon'$
\vspace{0.1cm}
\State {\textit{\#\# \quad Learning}}
\For{$ {\rm{Episode}} = 1$ \textbf{to} ${\rm{Max\_episode}}$}
\State Initialize a random process ${\mathcal{N}}$ for contract design exploration
\For{${\rm{Step}} = 1$ \textbf{to} ${\rm{Max\_step}}$}
\State Observe the current environment ${\bm{e}}_t$
\State {\textbf{Set ${\bm{c}}_t^N$ as Gaussian noise. Generate contract design ${\bm{c}}_t^0$ by denoising ${\bm{c}}_t^N$ using ${\bm{\varepsilon}}_{\theta}$ according to~\eqref{denoise}}}
\State Add the exploration noise $\epsilon$ to ${\bm{c}}_t^0$
\State Execute contract design ${\bm{c}}_t^0$ and observe reward $U_{SIR}$ that is defined as~\eqref{SIR}
\State Store the record $({\bm{e}}_t,{\bm{c}}_t^0,r_t)$ in replay buffer $R$
\State Sample a random minibatch of $N_b$ records $\left( {\bm{e}}_i,{\bm{c}}_i,r_i\right) $ from $R$
\State Set $y_i = r_i + \gamma {Q'}_{\upsilon'}\left( {\bm{e}}_{i},{\bm{c}'}_t^0\right) $, where ${\bm{c}'}_t^0$ is obtained using ${\bm{\varepsilon}'}_{\theta'}$
\State Update the {\textit{contract quality network}} by minimizing the loss:
\State \quad ${\mathcal{L}} = \frac{1}{N_b} \sum_i \left( y_i - Q_\upsilon\left( {\bm{e}}_i,{\bm{c}}_i\right) \right) ^2$
\State Update the {\textit{contract generation network}} by computing the policy gradient:
\State \quad $\nabla_{\theta}{\bm{\varepsilon}}_\theta \approx \frac{1}{N_b} \sum_i \nabla_{\bm{c}^0} Q_{\upsilon}\left( {\bm{e}},\bm{c}^0\right)|_{{\bm{e}}={\bm{e}}_i} \nabla_{\theta} {\bm{\varepsilon}}_\theta|{{\bm{e}}_i}$
\State Update the target networks:
\State \quad ${\theta'} \leftarrow \tau {\theta} + (1 - \tau) {\theta'}$,
\State \quad $\upsilon' \leftarrow \tau \upsilon + (1 - \tau) \upsilon'$
\EndFor
\EndFor
\State \Return The trained {\textit{contract generation network}} ${\bm{\varepsilon}}_\theta$
\end{algorithmic}
\hspace*{0.02in} {\bf{\textit{Inference Phase:}}}
\begin{algorithmic}[1]
\State Input the environment vector ${\bm e}$
\State Generate the optimal contract design ${\bm{c}}^0$ by denoising Gaussian noise using ${\bm{\varepsilon}}_\theta$ according to~\eqref{denoise}
\State \Return The optimal contract design ${\bm c}^0$
\end{algorithmic}
}
\end{algorithm}
}

\subsection{Performance Analysis}\label{fea}
After obtaining the semantic information according to the scheme in Section~\ref{S4A}, the user transmits the semantic information to other users in the form of D2D full-duplex communications with the contract incentive. Nevertheless, the wireless medium is subject to various impairments such as small-scale fading, self-interference, co-channel interference and noise, which inevitably degrade the achievable rate and BEP. To obtain the closed-form expressions with parameters capable of describing the above phenomenon in the semantic information D2D full-duplex transmission, we first perform the following performance analysis.
\subsubsection{PDF and CDF expressions}
To obtain the closed-form expressions of the achievable rate and the BEP of the D2D full-duplex communication links, we first derive the PDF expression of the SINR.
\begin{lemma}\label{lemma1}
The closed-form PDF and CDF expressions of the SINR can be derived as
\begin{align}\label{PDFfinal}
&{f_{{\gamma _k}}}\left( x \right) = \frac{{\alpha {x^{ - 1}}}}{{2\Gamma \left( N \right)\Gamma \left( \mu  \right)}}
\notag\\&\times \!\!
H_{1,0:1,1;1,1}^{0,0:0,1;1,1}\!\!\left(\!\!\!\!\!\! {\left. {\begin{array}{*{20}{c}}
{{{\left(\! {\frac{{\varphi xD_{jk}^{{\beta _k}}}}{{\beta {P_j}}}} \!\right)}^{\frac{{ - \alpha }}{2}}}} \\ 
{\frac{\varphi }{{{\eta _k}{P_I}}}} 
\end{array}} \!\!\!\right|\!\!\!\!\!\begin{array}{*{20}{c}}
{\left( {0;\! - \frac{\alpha }{2},1} \right)\!\!:\!\!\left(\! {1 - \mu ,1} \right)\!;\!\left( {1,1} \right)} \\ 
{ - :\left( {1,\frac{\alpha }{2}} \right);\left( {N,1} \right)} 
\end{array}} \!\!\!\!\!\right)\!,
\end{align}
and
\begin{align}\label{CDFfinal}
&	{F_{{\gamma _k}}}\left( x \right) = \frac{\alpha }{{2\Gamma \left( N \right)\Gamma \left( \mu  \right)}}
\notag\\&\times\!\!
H_{1,0:1,2;1,1}^{0,0:1,1;1,1}\!\!\left(\!\!\!\!\!\! {\left. {\begin{array}{*{20}{c}}
{{{\left(\!\! {\frac{{\beta {P_j}}}{{\varphi xD_{jk}^{{\beta _k}}}}} \!\!\right)}^{\frac{\alpha }{2}}}} \\ 
{\frac{\varphi }{{{\eta _k}{P_I}}}} 
\end{array}} \!\!\!\!\right|\!\!\!\!\!\begin{array}{*{20}{c}}
{\left( {0; - \frac{\alpha }{2},1} \right)\!\!:\!\!\left( {1 - \mu ,1} \right)\!\left( {1,\frac{\alpha }{2}} \right)\!;\!\left( {1,1} \right)} \\ 
{ - :\left( {0,\frac{\alpha }{2}} \right)\left( {1,\frac{\alpha }{2}} \right);\left( {N,1} \right)} 
\end{array}} \!\!\!\!\!\right)\!,
\end{align}
where $H_{ \cdot  \cdot }^{ \cdot  \cdot }\left( { \cdot \left|  \cdot  \right.} \right)$ is the Multivariate Fox's $H$-function \cite[eq. (A-1)]{mathai2009h}.
\begin{IEEEproof}
Please refer to Appendix~\ref{A1}.
\end{IEEEproof}
\end{lemma}
\subsubsection{Approximation Analysis}
\begin{figure}[t]
	\centering
	\includegraphics[width=0.4\textwidth]{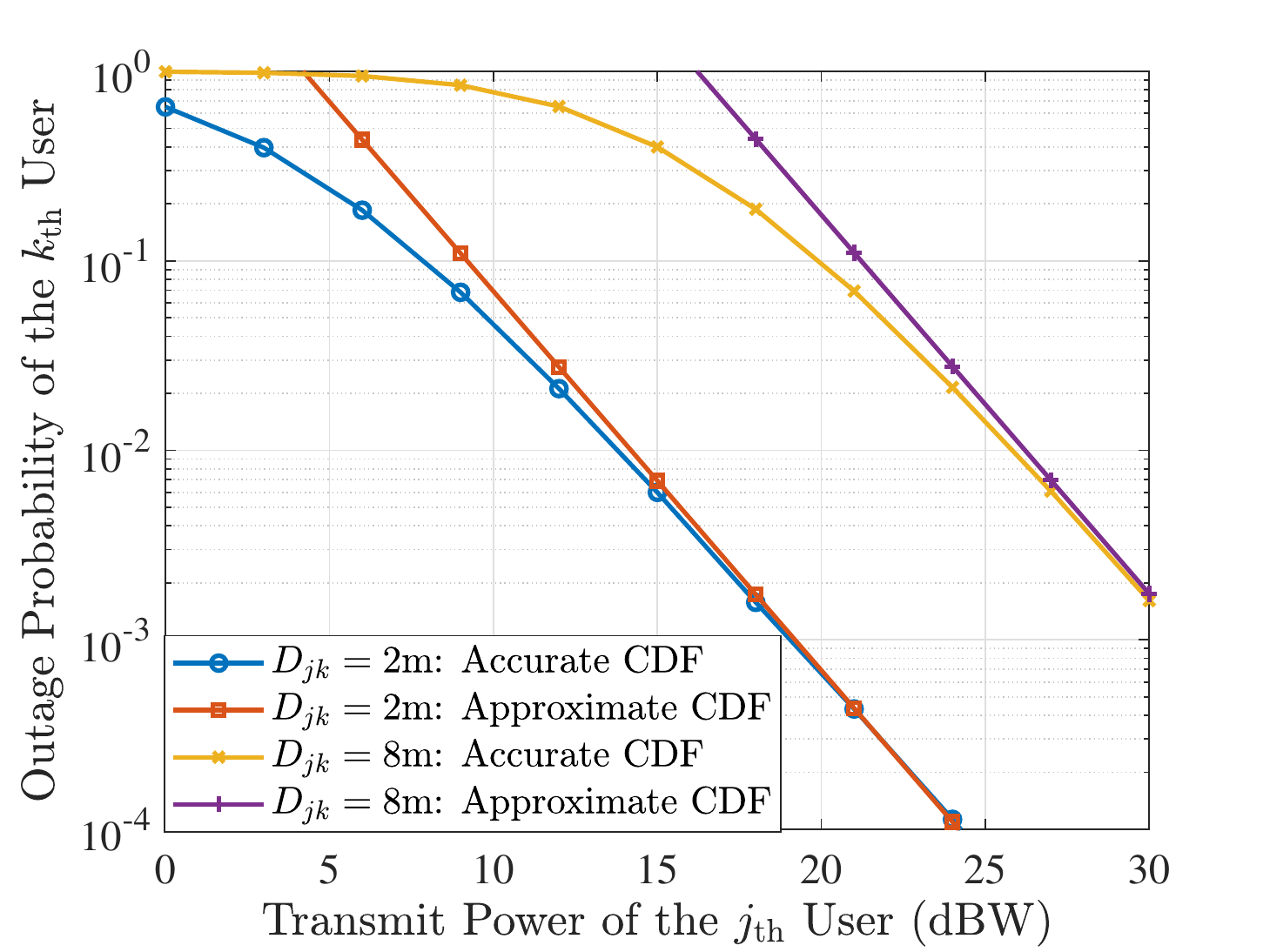}%
	\caption{The outage probability of the $k^{\rm th}$ user versus the transmit power of the $j^{\rm th}$ user under different transmit distance, with $\beta_{k} = 2$, $\alpha = 2$, $\mu = 2$, $P_k = 0$ ${\rm dBW}$, $P_I = 1$ ${\rm dBW}$, ${\eta _{k}} = 0.2$, ${\upsilon _k}=0.2$, and $N = 2$.}
	\label{opfigure}
\end{figure}

To simplify the mathematical tractability of the derived PDF and CDF expressions, we further study a practical special scenario, namely, the interference-limited case. Specifically, we consider that the noise is negligible, i.e., $\sigma_N^2 = 0$, and the user's HMD is equipped with an efficient self-interference cancellation technique, i.e., $\upsilon = 0$. Then, we can re-write the SNR as
\begin{equation}\label{SINRR}
{\gamma _{k'}} = \frac{{{P_j}D_{jk}^{ - {\beta _k}}{{\left| {{h_{jk}}} \right|}^2}}}{{{P_I}\sum\limits_{i = 1}^N {{{\left| {g_{k}} \right|}^2}}}}.
\end{equation}
The simplified PDF and CDF expressions can be derived.
\begin{lemma}\label{lemma2}
We derive the approximated PDF and CDF expressions when $\sigma_N^2 = 0$ and $\upsilon = 0$ as
\begin{align}\label{PDFfinal2}
	{f_{{\gamma _k}}}\left( x \right)& = \frac{{\alpha {x^{\frac{{\alpha \mu }}{2} - 1}}{{\left( {{\eta _k}{P_I}} \right)}^{\frac{{\alpha \mu }}{2}}}}}{{2{{\left( {{P_j}D_{jk}^{ - {\beta _k}}\beta } \right)}^{\frac{{\alpha \mu }}{2}}}\Gamma \left( N \right)\Gamma \left( \mu  \right)}}\notag\\&\times\!\!
H_{{\text{1}},{\text{1}}}^{{\text{1}},{\text{1}}}\!\left(\! {\left. {{{\left( {\frac{{\beta {P_j}D_{jk}^{ - {\beta _k}}}}{{{\eta _k}{P_I}x}}} \right)}^{\frac{\alpha }{2}}}} \right|\!\!\begin{array}{*{20}{c}}
		{\left( {1;1} \right)} \\ 
		{\left( {\frac{{\alpha \mu }}{2} + N;\frac{\alpha }{2}} \right)} 
\end{array}} \!\right),
\end{align}
and
\begin{align}\label{CDFfinal2}
&{F_{{\gamma _k}}}\left( x \right) = \frac{{\alpha {{\left( {y{\eta _k}{P_I}} \right)}^{\frac{{\alpha \mu }}{2}}}}}{{2{{\left( {{P_j}D_{jk}^{ - {\beta _k}}\beta } \right)}^{\frac{{\alpha \mu }}{2}}}\Gamma \left( N \right)\Gamma \left( \mu  \right)}}
\notag\\&\times\!\!
H_{{\text{2}},{\text{2}}}^{{\text{2}},{\text{1}}}\!\left(\!\! {\left. {{{\left(\! {\frac{{\beta {P_j}D_{jk}^{ - {\beta _k}}}}{{{\eta _k}{P_I}y}}} \!\right)}^{\frac{\alpha }{2}}}} \right|\!\!\!\begin{array}{*{20}{c}}
		{\left( {0;1} \right)\left( {\frac{{\alpha \mu }}{2} + 1;\frac{\alpha }{2}} \right)} \\ 
		{\left( {\frac{{\alpha \mu }}{2} + N;\frac{\alpha }{2}} \right)\left( {\frac{{\alpha \mu }}{2};\frac{\alpha }{2}} \right)} 
\end{array}} \!\!\right).
\end{align}
\begin{IEEEproof}
Please refer to Appendix~\ref{B1}.
\end{IEEEproof}
\end{lemma}
With the aid of Lemma~\ref{lemma2}, the performance analysis of a full-duplex system operating in an interference-limited scenario can be simplified. In particular, the outage probability (OP) can be expressed as the probability that the received SNR falls below a given outage threshold $\gamma^{\rm th}$, which can be mathematically represented as $P_O = {\mathbb P} \left( \gamma_k < \gamma^{\rm th}\right) =F{\gamma_k}\left( \gamma^{\rm th}\right)$. To evaluate the OP, we can utilize \eqref{CDFfinal2}. Additionally, when the transmit power of the $j{\rm th}$ user is high, i.e., $P_{j}\to \infty$, the OP can be further simplified by calculating the residue at the nearest pole to the Mellin-Barnes integral contour of the Meijer's $G$-function~\cite{gradshteyn2007}
\begin{equation}
	P_O \approx  \frac{{{{\left( {{\gamma _{th}}{\eta _k}{P_I}} \right)}^{\frac{{\alpha \mu }}{2}}}}}{{\mu {{\left( {{P_j}D_{jk}^{ - {\beta _k}}\beta } \right)}^{\frac{{\alpha \mu }}{2}}}\Gamma \left( N \right)\Gamma \left( \mu  \right)}}\Gamma \left( {\frac{{\alpha \mu }}{2} + N} \right).
\end{equation}
Figure~\ref{opfigure} shows the OP of the $k^{\rm th}$ user versus the transmit power of the $j^{\rm th}$ user under different transmit distance. As shown in the figure, the OP increases as the transmission distance increases. Moreover, in the high $P_{j}$ region, our derived approximate expression yields the result close to that from the exact expression, thereby validating the accuracy of our approximate analysis.

\subsubsection{Achievable Rate in Full-duplex Communications}
The ergodic achievable rate (or Shannon capacity) for the transmitter, which is known to be the maximum data rate that the channel can support (per Hz), is defined as
\begin{equation}\label{fdeklahjf}
R_{\rm k} = \int_0^\infty  {{{\log }_2}} (1 + x ){f_{\gamma_k} }(x ){\rm{d}}x.
\end{equation}
Then, with the help of Lemma~\ref{lemma1}, the achievable rate can be obtained in the following theorem.
\begin{them}\label{Theorem1}
The achievable rate of the $k^{\rm th}$ user, i.e., ${R_{\rm{k}}}$, can be derived in closed-form as
\begin{align} \label{ccfinal}
&{R_{\rm{k}}} = \frac{{{\Gamma ^{ - 1}}\left( \mu  \right)\alpha }}{{\ln 4\Gamma \left( N \right)}}
\notag\\&\times\!\!
H_{1,0:3,3;1,1}^{0,0:1,3;1,1}\!\!\!\left(\!\!\!\!\!\! {\left. {\begin{array}{*{20}{c}}
{{{\left(\!\! {\frac{{\beta {P_j}}}{{\varphi D_{jk}^{{\beta _k}}}}} \!\!\right)}^{\frac{\alpha }{2}}}} \\ 
{\frac{\varphi }{{{\eta _k}{P_I}}}} 
\end{array}} \!\!\!\!\right|\!\!\!\!\!\begin{array}{*{20}{c}}
{\left( {0;\!- \frac{\alpha }{2},\!1} \right)\!\!:\!\!\left(\!{1 \!-\! \mu ,\!1} \right)\!\!\left( {1,\frac{\alpha }{2}} \right)\!\!\left( {1,\frac{\alpha }{2}} \right)\!;\!\left( {1,\!1} \right)} \\ 
{ - :\left( {1,\frac{\alpha }{2}} \right)\left( {0,\frac{\alpha }{2}} \right)\left( {1,\frac{\alpha }{2}} \right);\left( {N,1} \right)} 
\end{array}} \!\!\!\!\!\right)\!.
\end{align}
\begin{IEEEproof}
Please refer to Appendix~\ref{A2}
\end{IEEEproof}
\end{them}

\subsubsection{Bit Error Probability in Full-duplex Communications}
The average BEP for the $k^{\rm th}$ user under a variety of modulation formats is given by~\cite{tse2005fundamentals}
{\small \begin{equation}\label{BEREQUATION}
E_{\rm k} = \int_0^\infty  {\frac{{\Gamma \! \left( {{\tau _2},{\tau _1}\gamma } \right)}}{{2\Gamma \! \left( {{\tau _2}} \right)}}{f_{\gamma_k}}\left( \gamma  \right)d\gamma },
\end{equation}}\noindent
where $\Gamma \left( { \cdot , \cdot } \right)$ is the upper incomplete Gamma function \cite[eq. (8.350.2)]{gradshteyn2007}, $ {{{\Gamma \left( {{\tau _2},{\tau _1}\gamma } \right)}}/{{2\Gamma \left( {{\tau _2}} \right)}}} $ is the conditional bit-error probability, ${\tau _1} $ and ${\tau _2}$ are modulation-specific parameters that represents various modulation/detection combinations. Specifically, $\left\{ {{\tau _1} = 1,{\tau _2} = 0.5} \right\}$ is antipodal coherent binary phase-shift keying (BPSK), $\left\{ {{\tau _1} = 0.5,{\tau _2} = 0.5} \right\}$ denote orthogonal coherent binary frequency-shift keying (BFSK), $\left\{ {{\tau _1} = 0.5,{\tau _2} = 1} \right\}$ is orthogonal non-coherent BFSK, and $\left\{ {{\tau _1} = 1,{\tau _2} = 1} \right\}$ denote antipodal differentially coherent BPSK (DPSK).

\begin{them}\label{Theorem2}
The closed-form expression of the BEP of the $k^{\rm th}$ user, i.e., $E_k$, is derived as
\begin{align}\label{berfinal}
&	{E_k} = \frac{{\alpha {\tau _1}^{\frac{\alpha }{2}{t_1}}{\Gamma ^{ - 1}}\left( {{\tau _2}} \right)}}{{4\Gamma \left( N \right)\Gamma \left( \mu  \right)}}
\notag\\&\times\!\!
H_{1,1:2,2;1,1}^{0,1:1,1;1,1}\!\!\left(\!\!\!\!\!\! {\left. {\begin{array}{*{20}{c}}
{{{\left(\! {\frac{{\beta {P_j}}}{{\varphi D_{jk}^{{\beta _k}}}}} \!\right)}^{\frac{\alpha }{2}}}} \\ 
{\frac{\varphi }{{{\eta _k}{P_I}}}} 
\end{array}} \!\!\!\!\right|\!\!\!\!\begin{array}{*{20}{c}}
{\left( {1; - \frac{\alpha }{2},1} \right)\!\!:\!\left( {1\! - \!\mu ,1} \right)\left( {1,\frac{\alpha }{2}} \right)\!;\!\left( {1,1} \right)} \\ 
{\left( {1;\frac{\alpha }{2}, - 1} \right)\!:\!\left( {0,\frac{\alpha }{2}} \right)\left( {1,\frac{\alpha }{2}} \right);\left( {N,1} \right)} 
\end{array}} \!\!\!\!\!\right).
\end{align}
\begin{IEEEproof}
Please refer to Appendix~\ref{A3}
\end{IEEEproof}
\end{them}
\begin{rem}
The above derived performance indicator expressions contain the important channel parameters in a D2D full-duplex wireless communications system. Specifically, parameters that can respond to large-scale fading, small-scale fading, self-interference in duplex communications, and co-channel interference at arbitrary multipath are all included in \eqref{ccfinal} and \eqref{berfinal}. Detailed discussion is given in Section~\ref{per}. Furthermore, the generalization of the small-scale fading model, i.e., $\alpha$-$\mu$ fading channel model, allows us to use the derived performance metrics in the analysis of many systems with various kinds of fading conditions.
\end{rem}

Thus, we complete all the processes of the framework. Specifically, we use the semantic encoder in Section~\ref{S4A} to extract interest point and descriptors. The computation task performed at the $k^{\rm th}$ user could be the free-space information detection, which can be obtained by using the state-of-art anomaly detection approach named JSR-Net~\cite{vojir2021road}. To motivate the semantic information sharing, we propose a contract theoretic mechanism to maximize the utility of SIR while ensuring that the utility of SIP is larger than the threshold of participation in sharing. In Section~\ref{ser2}, we propose the AI-generated contract algorithm based on the diffusion model to solve for the optimal contract design. For the D2D full-duplex wireless transmission, the performance analysis is comprehensively given in Section~\ref{fea}. After SIRs receive the free-space information, interest points, and descriptors, the semantic matching scheme can be used to help the users obtain the free-space information in their own view images by matching instead of performing computation tasks again.

\section{Numerical Results}\label{S6}
According to our core contributions, this section aims to answer the following research questions via experiments:
\begin{enumerate}
\item[{\textbf{Q1)}}] Can the proposed full-duplex semantic communication framework realize information sharing among users effectively?
\item[{\textbf{Q2)}}] On the basis of {\bf{Q1}}, can the proposed incentive mechanism further boost the performance of information sharing, i.e., the utility of the SIR?
\item[{\textbf{Q3)}}] How do wireless channel and self-interference affect semantic information transmission during full-duplex D2D communication?
\end{enumerate}
The experimental platform for running our proposed algorithms is built on a generic Ubuntu 20.04 system with an AMD Ryzen Threadripper PRO 3975WX 32-Cores CPU and an NVIDIA RTX A5000 GPU. In each Monte Carlo simulation, we generated $10^6$ random realizations of the channel to ensure statistical significance.
\subsection{Effectiveness of semantic communication framework (for {\bf{Q1}})}
\begin{figure*}[t]
\centering
\includegraphics[width=0.96\textwidth]{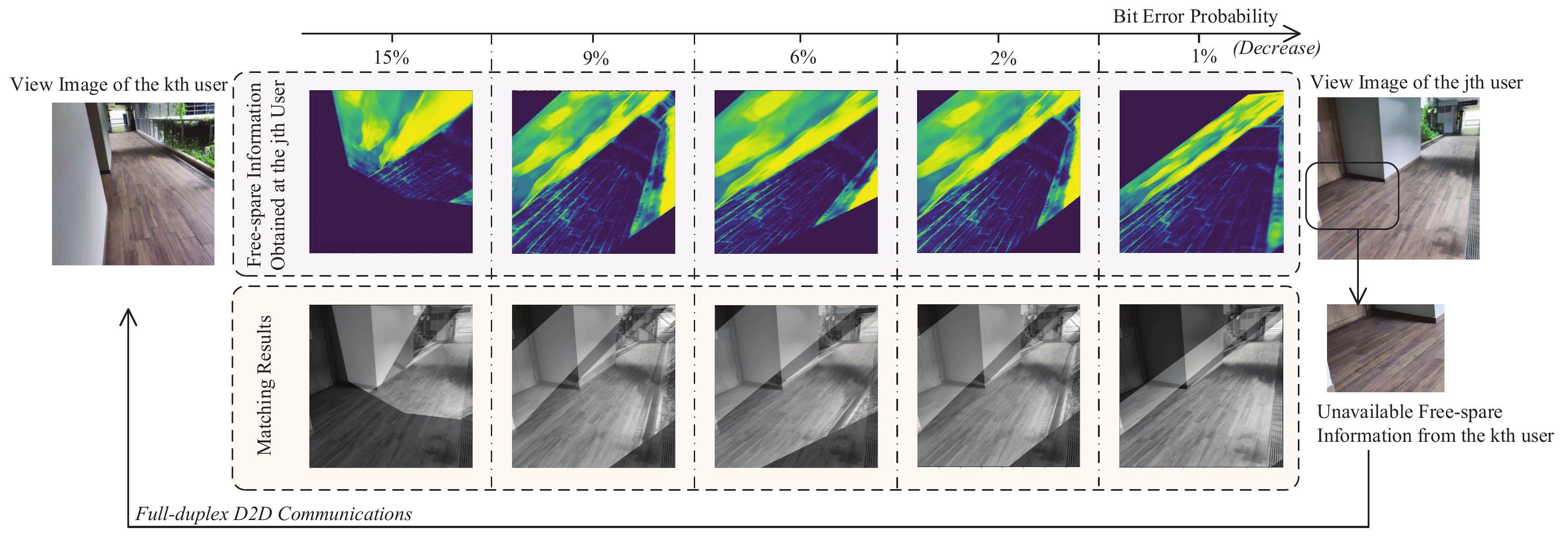}%
\caption{The impact of different wireless transmission BEPs on our proposed full-duplex D2D free-space information sharing mechanism. The semantic encoder extracts 150 interest points and corresponding descriptors in the image as the semantic information.}
\label{match}
\end{figure*}

First, we verify the effectiveness of the proposed D2D full-duplex semantic communications framework, and the results are shown in Fig.~\ref{match}. As can be seen from the experimental results, firstly, according to the interest points, descriptors, and free-space information transmitted by the ${k^{\rm th}}$ user, the ${j^{\rm th}}$ user can effectively compute his own free-space information, as shown by the images in the first row. Compared with the conventional approach, i.e., the ${j^{\rm th}}$ user computes the free-space information independently, this method consumes fewer resources since the ${j^{\rm th}}$ user only needs to compute the spatial correspondence information between view images and conduct a corresponding spatial rotation to obtain his own free-space information. Secondly, the results demonstrate that as the BEP decreases, the free-space information computed by the ${j^{\rm th}}$ user is more accurate, which finally generates a more accurate matching result, as presented by the images in the second row. Note that the values of BEP can be computed by using the environmental parameters and \eqref{berfinal}. This inspires us that the accuracy of semantic information during wireless transmission needs to be guaranteed in the free-space information sharing mechanism of D2D full-duplex communication systems.

It should be noted that due to the limitation of the viewing angle, some details, such as the corner marked by the black box in the ${j^{\rm th}}$ user's view, cannot be obtained using the semantic information extracted from the view image of the ${k^{\rm th}}$ user. However, this part of the information may be obtained through semantic information from users who have other different perspectives. Meanwhile, because full-duplex communication is taken into account here, the ${j^{\rm th}}$ user would feed back the missing information to the ${k^{\rm th}}$ user and other users, allowing the ${k^{\rm th}}$ user to obtain more free-space information, even if there is a blind spot. Such real-time information updating mechanism is also an advantage of our proposed D2D framework.
\subsection{Efficiency of the proposed incentive mechanism (for {\bf{Q2}})}\label{difftest}
\begin{figure}[!t]
\centering
\subfigure[Traning process, with diffusion step $N = 10$, batch size $N_b = 512$, discount factor $\gamma = 0.95$, soft target update parameter $\tau = 0.005$, exploration noise $\epsilon = 0.01$, contract generation network ${\bm{\varepsilon}}_\theta$ learning rate is $10^{-5}$, and contract quality network $Q_\upsilon$ learning rate is $10^{-5}$.]{\includegraphics[width=0.4\textwidth]{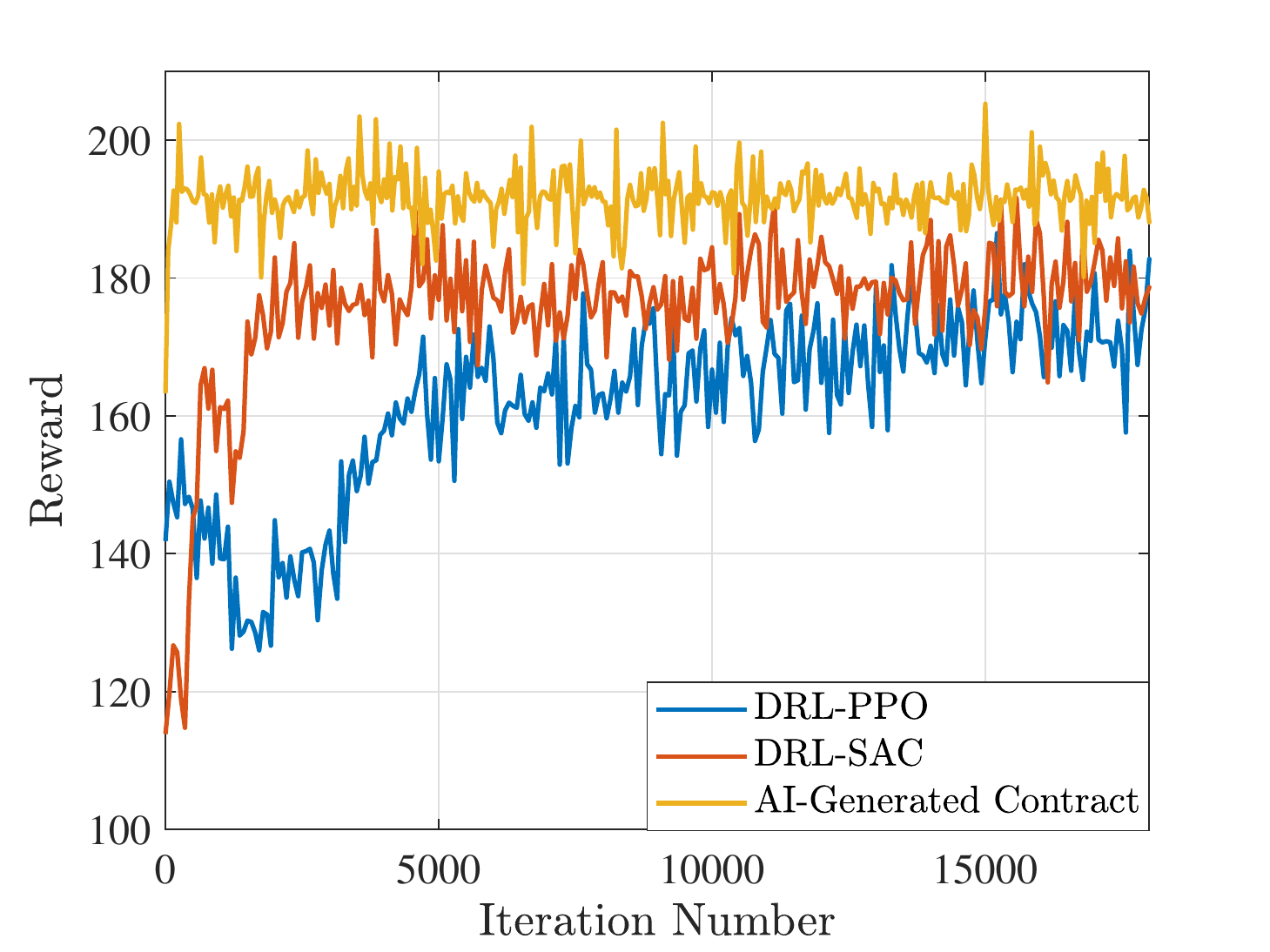}}
\subfigure[The designed contracts, with $c_{s} = 1000$, $c_p = 0.03$, $D_{jk} = 6$ m, $\alpha = 4$, $\mu = 4$, $\beta_{k} = 12.9$,  $\sigma _N^2 = \sigma _S^2 = 0.2$, $P_I = 3$ ${\rm dBW}$, ${\eta _{k}} = 0.3$, ${\upsilon _k}=0.2$, $B_{jk} = 10$ ${\rm MHz}$, and $N =3$.]{\includegraphics[width=0.4\textwidth]{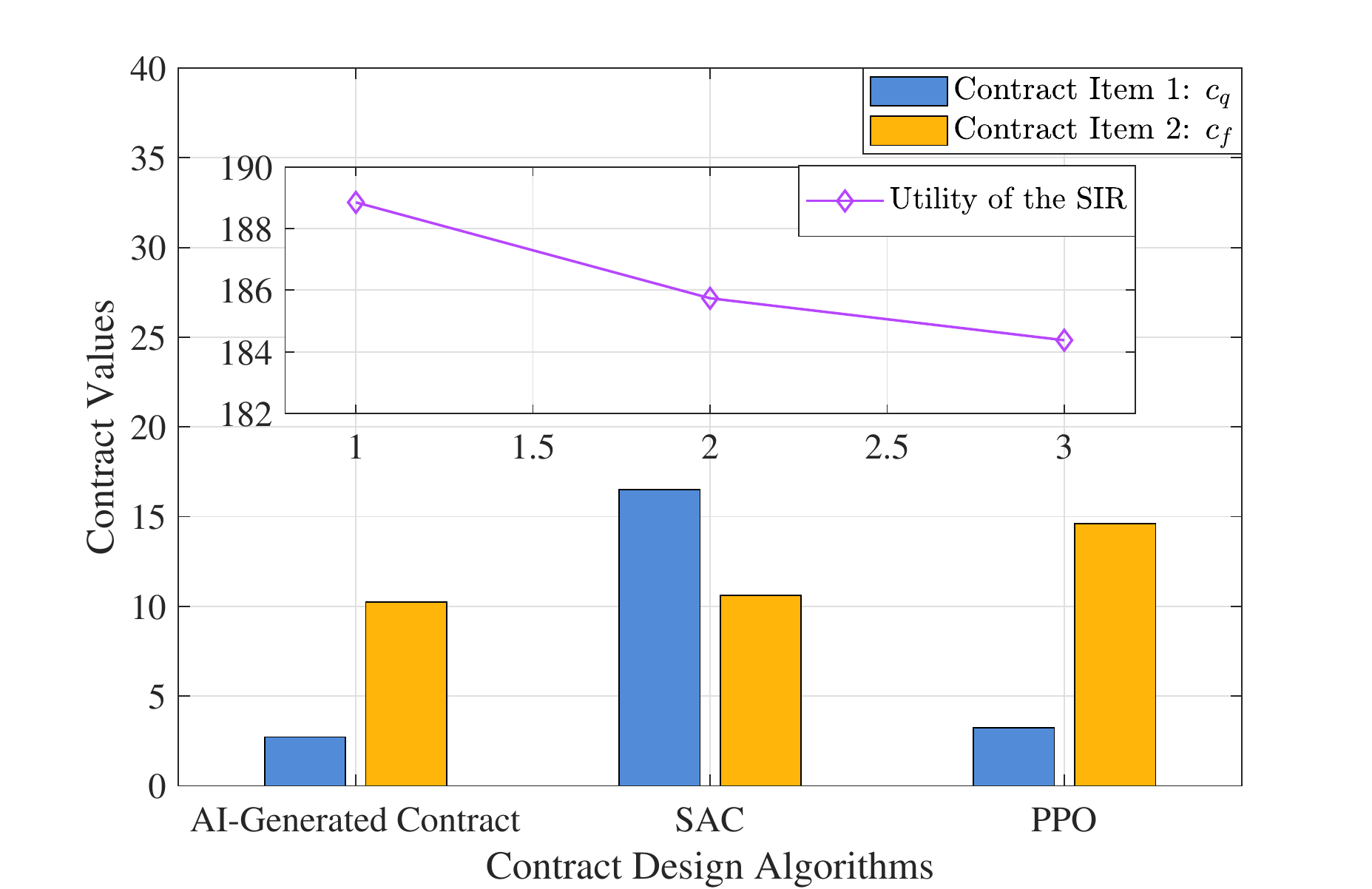}}
\caption{The effect of our proposed free-space sharing mechanism under different incentive design schemes, e.g., DRL-PPO, DRL-SAC, and AI-generated contract.}
\label{free}
\end{figure}

Second, we present the comparison between our proposed AI-generated contract algorithm and two conventional DRL algorithms, i.e., SAC and PPO. The training process shown in Fig~\ref{free} (a) reveals that PPO requires more iteration steps to converge. Although SAC can stabilize at a higher reward faster, both the convergence rate and the final reward value are inferior to those of the AI-generated contract algorithm. We posit two reasons for this.
	\begin{itemize}
		\item Our algorithm achieves better sampling quality. Diffusion models can generate higher quality samples by fine-tuning multiple times. For example, we set the diffusion step to 10. Since each fine-tuning gradually adjusts the model's output, the effect of uncertain and noise can be reduced and the sampling accuracy is improved.
		\item Our algorithm possesses better long-term dependence processing capability. While conventional neural network generation models consider only the input at the current time step, the diffusion model allows the model to generate samples with more time steps by fine-tuning multiple times, thereby having better long-term dependence processing capability.
	\end{itemize}
	Furthermore, we compare the trained models' ability for optimal contract design in Fig~\ref{free} (b). For a given environment state, we observe that the AI-generated contract algorithm can provide a contract design that achieves the SIR utility value of 189.1, which is greater than the 185.9 achieved by the SAC and the 184.3 achieved by the PPO.
\subsection{Influence of channel and self-interference (for {\bf{Q3}})}\label{per}
\begin{figure}[t]
\centering
\includegraphics[width=0.42\textwidth]{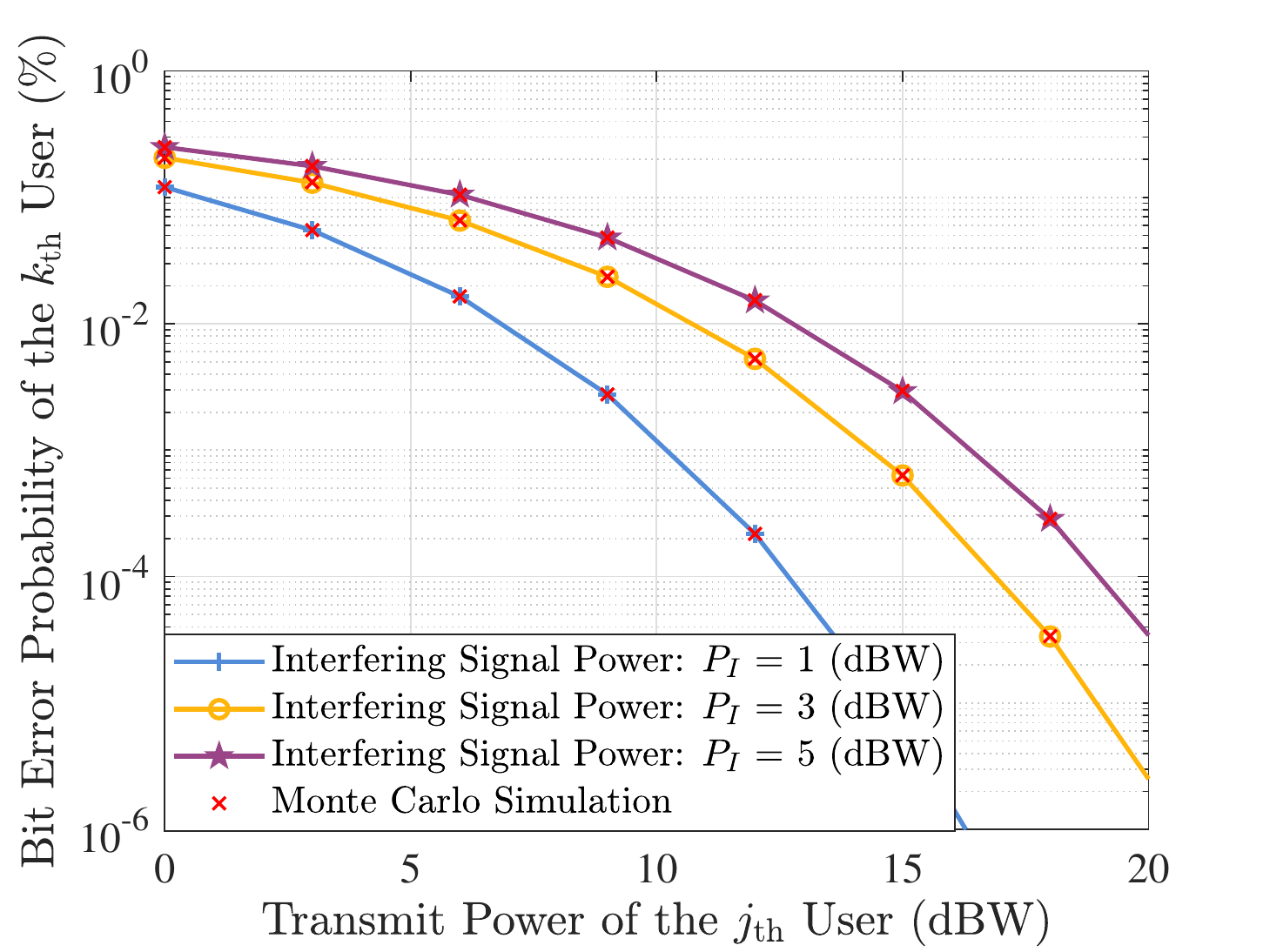}
\caption{The bit error probability of the $k^{\rm th}$ user versus the transmit power of the $j^{\rm th}$ user under different interfering signal power, with $D_{jk} = 5$ m, $\beta_{k} = 2$, $\alpha = 4$, $\mu = 5$, $\sigma _N^2 = \sigma _S^2 = 0.1$, $P_k = 20$ ${\rm dBW}$, $P_I = 1$ ${\rm dBW}$, ${\eta _{k}} = 0.2$, ${\upsilon _k}=0.2$, and $N = 2$.}
\label{BER1}
\end{figure}

Third, we analyze the impact of wireless channels and self-interference on D2D full-duplex semantic communications. The effect of the ${j^{\rm th}}$ user's (transmitter's) transmit power on the BEP of the ${k^{\rm th}}$ user (receiver) is investigated in Fig.~\ref{BER1}, during which the transmit power of ${k^{\rm th}}$ user is fixed to $20$ dBW. As the results show, as the transmit power of ${j^{\rm th}}$ user increases from $0$ to $20$ dBW, the BEP of ${k^{\rm th}}$ user decreases from about ${10^{ - 1}}$ to ${10^{ - 5}}$. The clear reason is that since the increase in transmit power makes the SINR higher and thus reduces the BEP. Furthermore, the increase in interfering signal power raises the BEP, and the degree of BEP increase goes up as the transmit power increases. When the transmit power is 0 dBW, for example, the BEPs corresponding to different interfering signal power are relatively close, whereas the BEP differences become larger when the transmit power reaches 20 dBW.

\begin{figure}[t]
\centering
\includegraphics[width=0.42\textwidth]{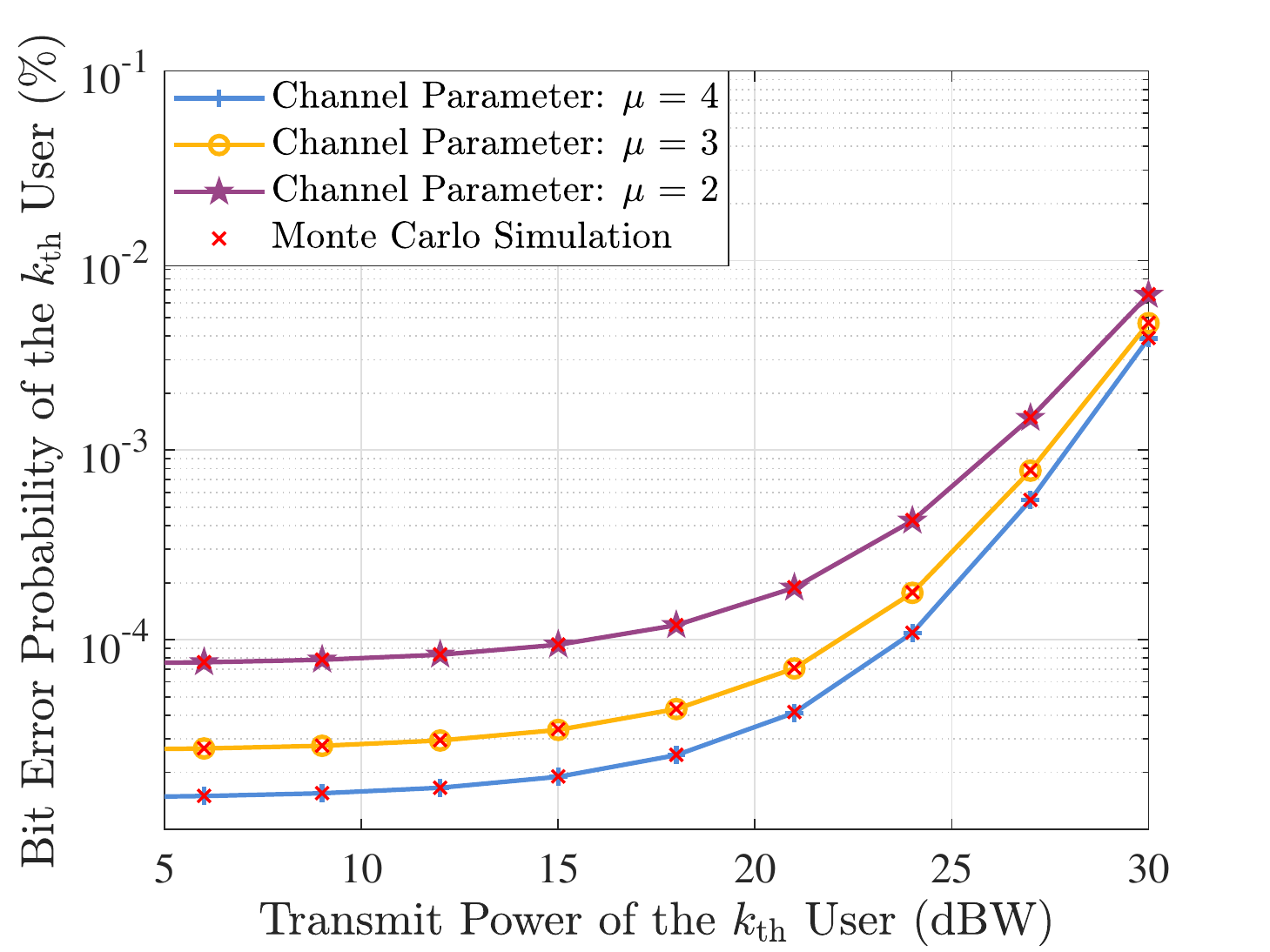}%
\caption{The bit error probability of the $k^{\rm th}$ user versus the transmit power of the $k^{\rm th}$ user under different channel parameters, with $D_{jk} = 10$ m, $\beta_{k} = 2$, $\alpha = 4$, $\mu = 4$, $\sigma _N^2 = \sigma _S^2 = 0.1$, $P_j = 20$ ${\rm dBW}$, $P_I = 1$ ${\rm dBW}$, ${\eta _{k}} = 0.2$, ${\upsilon _k}=0.2$, and $N = 2$.}
\label{BER2}
\end{figure}

Then, the influence of the ${k^{\rm th}}$ user's own transmit power on BEP is analyzed. The experimental results in Fig.~\ref{BER2} demonstrate that with the increase of transmit power, the BEP of ${k^{\rm th}}$ user increases. Taking the case of the channel parameter $\mu{\rm{ = }}2$ as an example, the BEP is smaller than ${10^{ - 4}}$ when the transmit power is 5 dBW, while it reaches nearly ${10^{ - 2}}$ as the transmit power arrives at 30 dBW. This can be interpreted by that the enhancement of transmit power would cause greater self-interference, which further affects the BEP. In addition, when the transmit power is fixed, the BEP decreases as the channel parameter increases, indicating that the severity of multipath effects directly impacts the ${k^{\rm th}}$ user's BEP.

\begin{figure}[t]
\centering
\includegraphics[width=0.42\textwidth]{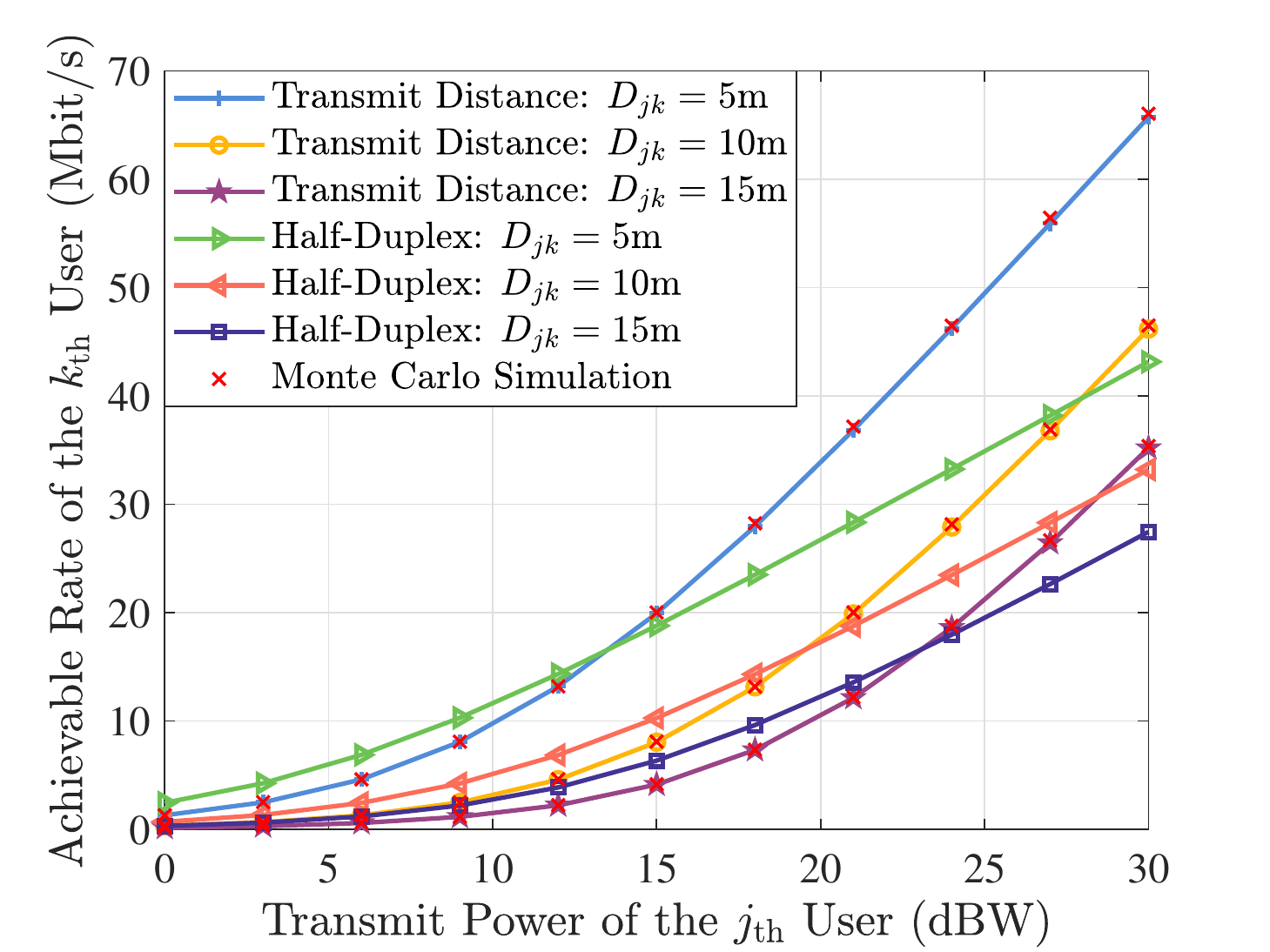}%
\caption{The achievable rate of the $k^{\rm th}$ user versus the transmit power of the $j^{\rm th}$ user under different transmit distance, with $\beta_{k} = 2$, $\alpha = 4$, $\mu = 5$, $\sigma _N^2 = \sigma _S^2 = 0.8$, $P_j = 10$ ${\rm dBW}$, $P_I = 1$ ${\rm dBW}$, ${\eta _{k}} = 0.2$, ${\upsilon _k}=0.2$, $B_{jk} = 10$ ${\rm MHz}$ and $N = 2$, under the full-duplex and half-duplex modes.}
\label{CC2}
\end{figure}

Next, under the condition of different distances between transmitter and receiver, the influence of transmit power on achievable rate is analyzed. From the results in Fig.~\ref{CC2}, one can see that the achievable rate increases with the boost of the transmit power. For instance, when the distance between the SIP and SIR is 5 meters, the achievable rate is approximately 1 Mbit/s at 0 dBW, but it reaches 50 Mbit/s when the transmit power is increased to 25 dBW. The reason is that increasing transmit power raises the SINR. Conversely, increasing the distance between the transmitter and the receiver would lead to a decrease in the SINR. Furthermore, we compare the performance of full-duplex and half-duplex (with time division multiple access scheme) modes. Our results demonstrate that when the transmit power of the SI, i.e., the $j^{\rm th}$user , is small, the achievable rate in full-duplex mode is lower than that in half-duplex mode due to the impact of strong self-interference. However, since the bandwidth can be shared, the increase in SIP's transmit power has a more significant impact on improving achievable rate in the full-duplex mode.

\begin{figure}[t]
\centering
\includegraphics[width=0.42\textwidth]{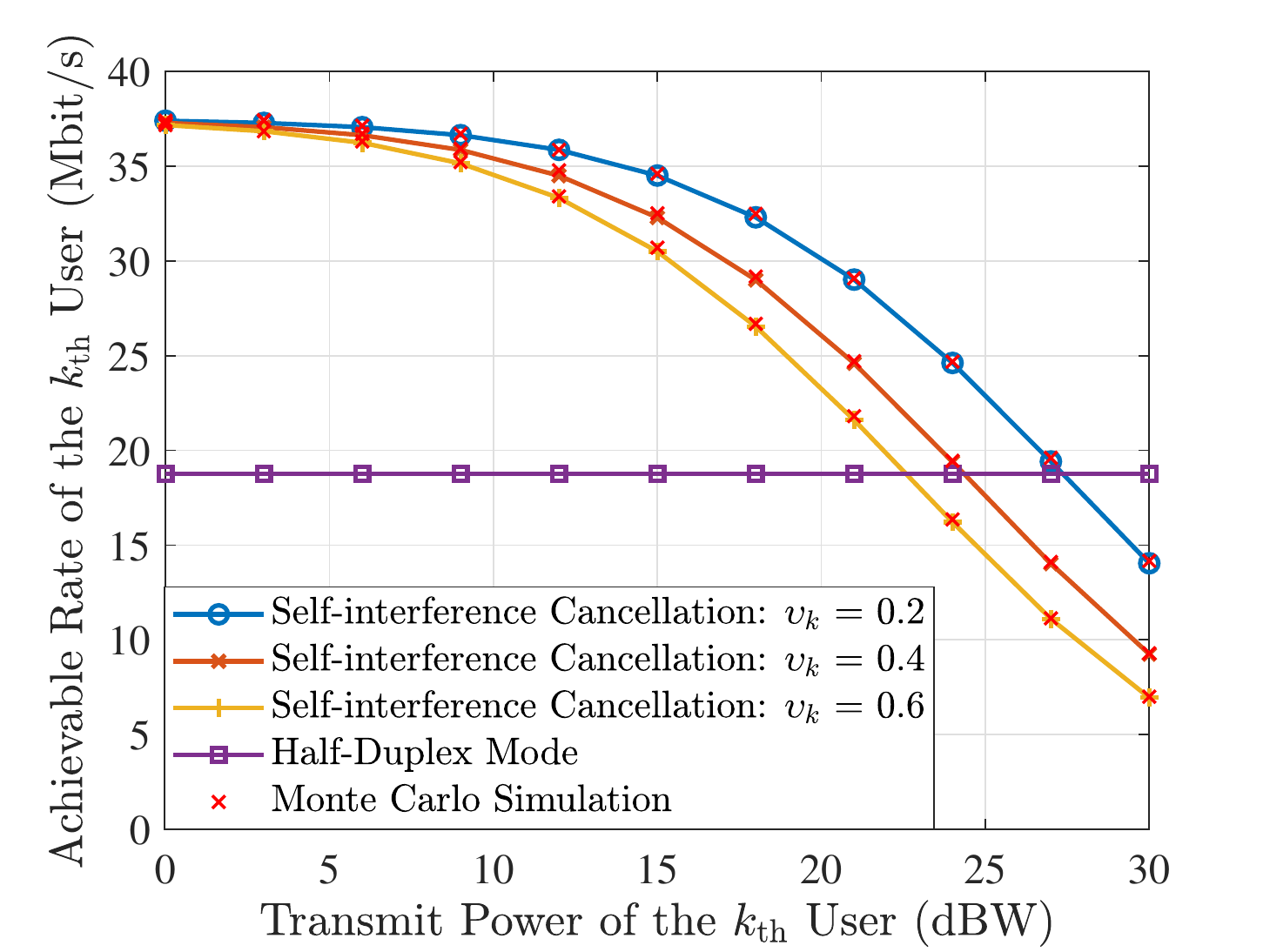}%
\caption{The achievable rate of the $k^{\rm th}$ user versus the transmit power of the $k^{\rm th}$ user under different self-interference cancellation, with $D_{jk} = 5$ m, $\beta_{k} = 2$, $\alpha = 4$, $\mu = 5$, $\sigma _N^2 = \sigma _S^2 = 0.1$, $P_j = 10$ ${\rm dBW}$, $P_I = 1$ ${\rm dBW}$, ${\eta _{k}} = 0.2$, $B_{jk} = 10$ ${\rm MHz}$ and $N = 2$, under the full-duplex and half-duplex modes.}
\label{CC1}
\end{figure}

We proceed to investigate the impact of the ${k^{\rm th}}$ user's own transmit power on the achievable rate and the results are shown in Fig.~\ref{CC1}. Specifically, when ${v_k}$ is fixed, increasing the transmit power of the ${k^{\rm th}}$ user would cause a decrease to the achievable rate, since the power enhancement introduce greater self-interference. From another point of view, it can be seen that improving the interference cancellation technique, i.e., decreasing ${v_k}$, can enhance the achievable rate. In the half-duplex mode, the received achievable rate at the SIR's side is not affected by its own transmit power, as users do not transmit and receive data simultaneously. By comparing the full-duplex and half-duplex modes, we conclude that the half-duplex mode could achieve a higher achievable rate than the full-duplex mode when the user's own transmit power is higher and the self-interference cancellation technique is poor.

\section{Conclusion}\label{S7}
We studied a D2D full-duplex communications system in which a free-space information-sharing mechanism was proposed to avoid redundant computation tasks. Specifically, the semantic encoder in the SIP extracts semantic information, i.e., interest points and descriptors, from the view image. Computation results and semantic information are then transmitted to SIRs in full-duplex communication mode. We modeled the wireless channel using a generalized small-scale fading model and comprehensively analyzed the effects of self-interference and multiple co-channel interference signals. The closed-form achievable rate and BEP were derived. After the SIR receives the semantic information, it can perform semantic matching to save energy for redundant computation tasks. We proposed a contract-based mechanism to incentivize semantic information sharing. A novel diffusion model-based AI-generated contract design algorithm was proposed, achieving higher SIR utility than DRL-based algorithms. For future work, our proposed semantic information-sharing framework can be further enhanced by considering more accurate self-interference modeling. The proposed diffusion model-based algorithm holds promise for solving various optimization problems in wireless networks.

\begin{appendices}
\section{Proof for Lemma~\ref{lemma1}}\label{A1}
\renewcommand{\theequation}{A-\arabic{equation}}
\setcounter{equation}{0}
\subsection{Proof of the PDF expression}
Because $ {{{\left| {g_{k}} \right|}^2}} \sim {\rm Rayleigh} \left( {\eta_{k}}\right)$ and the sum of $N$ i.i.d. Rayleigh-fading signals have a Nakagami-$m$ distributed signal amplitude with $m = N$, the PDF and CDF expressions of $ U \triangleq {\sum\limits_{i = 1}^N {{{\left| {g_{k}} \right|}^2}} } $ can be written as follows \cite{nakagami1960m}. 
\begin{equation}\label{PDFnakagemi}
f_{U}\left( x \right) = \frac{{{x^{N - 1}}}}{{\eta _{k}^N\Gamma \left( N \right)}}\exp \left( { - \frac{x}{{{\eta _{k}}}}} \right),
\end{equation}
%and
%\begin{equation}\label{CDFnakagemi}
%F_{U}\left( x \right) = \frac{{\gamma \left( {N,\frac{x}{{{\eta _{k}}}}} \right)}}{{\Gamma \left( N \right)}},
%\end{equation}
where $ {{\eta_{k}}}={{{\mathbb E}\left[ g_{k}^2 \right]}}$, and ${\mathbb E}\left[ \cdot \right]$ denotes expectation.
%, and $\gamma\left(\cdot,\cdot\right)$ is the incomplete Gamma function defined in \cite[eq. (8.350.1)]{gradshteyn2007}.

Let $ V \triangleq {P_I}U + {\upsilon_k}{P_k}\sigma _S^2 + \sigma _N^2 $ and $\varphi \triangleq {{\upsilon_k}{P_k}\sigma _S^2 + \sigma _N^2} $. Then, the PDF of $V$ can be derived as
\begin{equation}\label{PDFV}
{f_V}\left( y \right) = \frac{1}{{{P_I}}}{f_U}\left( {\frac{{y - \varphi}}{{{P_I}}}} \right).
\end{equation}
Substituting \eqref{PDFnakagemi} into \eqref{PDFV}, we have
\begin{equation}
{f_V}\left( y \right) = \frac{{{{\left( {y - \varphi} \right)}^{N - 1}}}}{{{{\left( {{P_I}{\eta _k}} \right)}^N}\Gamma \left( N \right)}}\exp \left( { - \frac{{y - \varphi}}{{{\eta _k}{P_I}}}} \right).
\end{equation}
Let $ W \triangleq {P_j}D_{jk}^{ - {\beta _k}}{\left| {{h_{jk}}} \right|^2} $. We can obtain that
\begin{equation}\label{PDFW}
{f_W}\left( w \right) = \frac{1}{{{P_j}D_{jk}^{ - {\beta _k}}}}{f_{{{\left| {{h_{jk}}} \right|}^2}}}\left( {\frac{w}{{{P_j}D_{jk}^{ - {\beta _k}}}}} \right)
\end{equation}
By substituting \eqref{PDFHJK} into \eqref{PDFW},the PDF of $W$ can be derived as
\begin{equation}
{f_W}\left( w \right) = \frac{{\alpha {{\left( w \right)}^{\frac{{\alpha \mu }}{2} - 1}}}}{{2{{\left( {{P_j}D_{jk}^{ - {\beta _k}}\beta } \right)}^{\frac{{\alpha \mu }}{2}}}\Gamma \left( \mu  \right)}}\exp \left( { - {{\left( {\frac{w}{{\beta {P_j}D_{jk}^{ - {\beta _k}}}}} \right)}^{\frac{\alpha }{2}}}} \right).
\end{equation}
Because $ {\gamma _k} = \frac{W}{V} $, we can derive the PDF of ${\gamma _k}$ as
\begin{equation}\label{PDFGA}
{f_{{\gamma _k}}}\left( x \right) = \int_0^\infty  {y{f_W}\left( {xy} \right){f_V}\left( y \right){\rm{d}}y} 
\end{equation}
Substituting \eqref{PDFW} and \eqref{PDFV} into \eqref{PDFGA}, we have
\begin{equation}\label{pdfd}
{f_{{\gamma _k}}}\left( x \right) = \frac{{\alpha {x^{\frac{{\alpha \mu }}{2} - 1}}}}{{2{{\left( {{P_j}D_{jk}^{ - {\beta _k}}\beta } \right)}^{\frac{{\alpha \mu }}{2}}}{{\left( {{P_I}{\eta _k}} \right)}^N}\Gamma \left( N \right)\Gamma \left( \mu  \right)}}{I_1},
\end{equation}
where
\begin{align}\label{i1}
{I_1} = & \int_{\varphi }^\infty  {{y^{\frac{{\alpha \mu }}{2}}}{{\left( {y - \varphi} \right)}^{N - 1}}} \exp \left( { - \frac{{y - \varphi}}{{{\eta _k}{P_I}}}} \right)
\notag\\&
\times \exp \left( { - {{\left( {\frac{{xy}}{{\beta {P_j}D_{jk}^{ - {\beta _k}}}}} \right)}^{\frac{\alpha }{2}}}} \right){\rm{d}}y
\end{align}
To solve for the integral in ${I_1}$, we map the exponential function to the complex domain. According to \cite[eq. (01.03.07.0001.01)]{web} as a Mellin-Barnes integral
\begin{equation}\label{exp1}
\exp \left( { - {{\left( {\frac{{xy}}{{\beta {P_j}D_{jk}^{ - {\beta _k}}}}} \right)}^{\frac{\alpha }{2}}}} \right)=\int_{{\mathcal{L}_1}} {\frac{{\Gamma \left( {{s_1}} \right)}}{{2\pi i}}{{\left( {\frac{{xy}}{{\beta {P_j}D_{jk}^{ - {\beta _k}}}}} \right)}^{\frac{{ - {s_1}\alpha }}{2}}}{\rm{d}}{s_1}},
\end{equation}
where the integration path of  ${\cal L}_1$ goes from $s_1 -\infty j$ to $s_1 +\infty j$ and $s_1 \in \mathbb{R}$, and
\begin{align}\label{exp2}
\exp \left( { - \frac{{y - \varphi}}{{{\eta _k}{P_I}}}} \right)=\frac{1}{{2\pi i}}\int_{{\mathcal{L}_2}} {\Gamma \left( {{s_2}} \right)} {\left( {\frac{{y - \varphi}}{{{\eta _k}{P_I}}}} \right)^{ - {s_2}}}{\rm{d}}{s_2},
\end{align}
where the integration path of  ${\cal L}_2$ goes from $s_2 -\infty j$ to $s_2 +\infty j$ and $s_2 \in \mathbb{R}$.
Substituting \eqref{exp1} and \eqref{exp2} into \eqref{i1} and changing the order of integration, we can rewrite $I_1$ as
\begin{align}
{I_1} =& {\left( {\frac{{\text{1}}}{{{\text{2}}\pi i}}} \right)^2}\int_{{\mathcal{L}_1}} {\int_{{\mathcal{L}_2}} {\Gamma \left( {{s_1}} \right)\Gamma \left( {{s_2}} \right)} {{\left( {\frac{1}{{{\eta _k}{P_I}}}} \right)}^{ - {s_2}}}} 
\notag\\&\times
{\left( {\frac{x}{{\beta {P_j}D_{jk}^{ - {\beta _k}}}}} \right)^{\frac{{ - {s_1}\alpha }}{2}}}{I_2}{\rm{d}}{s_1}{\rm{d}}{s_2},
\end{align}
where
\begin{equation}
{I_2} = \int_\varphi ^\infty  {{y^{\frac{{\alpha \mu  - {s_1}\alpha }}{2}}}{{\left( {y - \varphi } \right)}^{N - 1 - {s_2}}}{\rm d}y} 
\end{equation}
With the help of~\cite[eq. (3.191.2)]{gradshteyn2007}, $I_2$ can be solved as
\begin{equation}\label{i2}
{I_2} = {\varphi ^{N - {s_2} + \frac{{\alpha \mu  - {s_1}\alpha }}{2}}}B\left( {{s_2} - N - \frac{{\alpha \mu  - {s_1}\alpha }}{2},N - {s_2}} \right),
\end{equation}
where $B(\cdot, \cdot)$ denotes the beta function \cite[eq. (8.38)]{gradshteyn2007}. Substituting \eqref{i1} and \eqref{i2} into \eqref{pdfd} and using \cite[eq. (8.384.1)]{gradshteyn2007}, we can obtain the PDF expression of ${\gamma _k}$ in the Mellin-Barnes integral as
%$  {s_1}> \mu  $, $ N >  {s_2}> 0 $, and 
\begin{align}
{f_{{\gamma _k}}}\left( x \right)& = \frac{{\alpha {x^{\frac{{\alpha \mu }}{2} - 1}}{\varphi ^{N + \frac{\mu }{2}\alpha }}}}{{2{{\left( {{P_j}D_{jk}^{ - {\beta _k}}\beta } \right)}^{\frac{{\alpha \mu }}{2}}}{{\left( {{P_I}{\eta _k}} \right)}^N}\Gamma \left( N \right)\Gamma \left( \mu  \right)}}{\left( {\frac{{\text{1}}}{{{\text{2}}\pi i}}} \right)^2}
\notag\\&\times
\int_{{\mathcal{L}_1}} {\int_{{\mathcal{L}_2}} {\frac{{\Gamma \left( {{s_1}} \right)\Gamma \left( {{s_2}} \right)\Gamma \left( {N - {s_2}} \right)}}{{\Gamma \left( {\frac{{{s_1} - \mu }}{2}\alpha } \right)\Gamma \left( {{s_2} - N + \frac{{{s_1} - \mu }}{2}\alpha } \right)}}} } 
\notag\\&\times
{\left( {\frac{\varphi }{{{\eta _k}{P_I}}}} \right)^{ - {s_2}}}{\left( {\frac{{\varphi x}}{{\beta {P_j}D_{jk}^{ - {\beta _k}}}}} \right)^{\frac{{ - {s_1}\alpha }}{2}}}{\rm{d}}{s_1}{\rm{d}}{s_2}
\end{align}
Let $ {s_1} \triangleq {t_1} + \mu  $ and $ {s_2} \triangleq N - {t_2} $. With the help of \cite[eq. (A-1)]{mathai2009h}, the PDF of ${\gamma _k}$ can be expressed in form of the Multivariate Fox's $H$-function as \eqref{PDFfinal} to complete the proof.
\subsection{Proof of the CDF expression}
With the help of the definition of CDF, we have 
\begin{equation}\label{CDFd}
{F_{{\gamma _k}}}\!\left( x  \right) 	= \int_0^x  \!{{f_{{\gamma _k}}}\left( y \right){\rm d}y}
\end{equation}
Substituting \eqref{PDFfinal} into \eqref{CDFd} and using \cite[eq. (A-1)]{mathai2009h}, we can re-write the CDF of $\gamma_k$ as
\begin{align}\label{tempcdf}
&	{F_{{\gamma _k}}}\left( x \right) = \frac{{\alpha {{\left( {{\text{2}}\pi i} \right)}^{ - 2}}}}{{2\Gamma \left( N \right)\Gamma \left( \mu  \right)}}\int_{{\mathcal{L}_1}} {\int_{{\mathcal{L}_2}} {\frac{{\Gamma \left( {{t_1} + \mu } \right)\Gamma \left( {{t_2}} \right)\Gamma \left( {N - {t_2}} \right)}}{{\Gamma \left( {\frac{\alpha }{2}{t_1} - {t_2}} \right)\Gamma \left( {\frac{\alpha }{2}{t_1}} \right)}}} } 
\notag\\&\times
{\left( {\frac{\varphi }{{\beta {P_j}D_{jk}^{ - {\beta _k}}}}} \right)^{\frac{{ - \alpha }}{2}{t_1}}}{\left( {\frac{\varphi }{{{\eta _k}{P_I}}}} \right)^{{t_2}}}{I_3}{\rm{d}}{t_1}{\rm{d}}{t_2},
\end{align}
where $I_3$ can be solved by using \cite[eq. (8.334.3)]{gradshteyn2007} as
\begin{equation}\label{I3}
{I_3} = \int_0^x {{y^{\frac{{ - \alpha }}{2}{t_1} - 1}}{\rm d}y} = \frac{{\Gamma \left( { - \frac{\alpha }{2}{t_1}} \right)}}{{\Gamma \left( {{\text{1}} - \frac{\alpha }{2}{t_1}} \right)}}{x^{\frac{{ - \alpha }}{2}{t_1}}}.
\end{equation}
By substituting $I_3$ into \eqref{tempcdf} and using \cite[eq. (A-1)]{mathai2009h}, the CDF of $\gamma_k$ can be derived as \eqref{CDFfinal}, which completes the proof.
\section{Proof for Lemma~\ref{lemma2}}\label{B1}
\renewcommand{\theequation}{B-\arabic{equation}}
\setcounter{equation}{0}
\subsection{Proof of the PDF expression}
With the help of \eqref{pdfd} and \eqref{exp1}, we can express the PDF of ${\gamma _{k'}}$ as
\begin{equation}\label{faef}
{f_{{\gamma _{k'}}}}\left( x \right) = \frac{{\alpha {x^{\frac{{\alpha \mu }}{2} - 1}}}}{{2{{\left( {{P_j}D_{jk}^{ - {\beta _k}}\beta } \right)}^{\frac{{\alpha \mu }}{2}}}{{\left( {{P_I}{\eta _k}} \right)}^N}\Gamma \left( N \right)\Gamma \left( \mu  \right)}}{I_B},
\end{equation}
where
\begin{align}
{I_B} = & \int_{{\mathcal{L}_1}} {\frac{{\Gamma \left( {{s_1}} \right)}}{{{\text{2}}\pi i}}{{\left( {\frac{x}{{\beta {P_j}D_{jk}^{ - {\beta _k}}}}} \right)}^{\frac{{ - {s_1}\alpha }}{2}}}} 
\notag\\&\times
\int_0^\infty  {{y^{\frac{{\alpha \mu }}{2} + N - 1 - \frac{{{s_1}\alpha }}{2}}}} \exp \left( { - \frac{y}{{{\eta _k}{P_I}}}} \right){\rm{d}}y{\rm{d}}{s_1}
\end{align}
Using \cite[eq. (3.351.3)]{gradshteyn2007}, we can solve the integration in $I_B$. Substituting $I_B$ into \eqref{faef} and using \cite[eq. (A-1)]{mathai2009h}, we derive the PDF as~\eqref{PDFfinal2}.
\subsection{Proof of the CDF expression}
According to the definition of CDF and \eqref{PDFfinal2}, we have 
\begin{align}\label{faekj}
&	{F_{{\gamma _k}}}\left( x \right) = \frac{{\alpha {{\left( {{\eta _k}{P_I}} \right)}^{\frac{{\alpha \mu }}{2}}}{\Gamma ^{ - 1}}\left( \mu  \right)}}{{2{{\left( {{P_j}D_{jk}^{ - {\beta _k}}\beta } \right)}^{\frac{{\alpha \mu }}{2}}}\Gamma \left( N \right)}}\int_{{\mathcal{L}_1}} {\Gamma \left( {\frac{{\alpha \left( {\mu  - {s_1}} \right)}}{2} + N} \right)} 
	\notag\\&\times
\frac{{\Gamma \left( {{s_1}} \right)}}{{{\text{2}}\pi i}}{\left( {\frac{{{\eta _k}{P_I}}}{{\beta {P_j}D_{jk}^{ - {\beta _k}}}}} \right)^{\frac{{ - {s_1}\alpha }}{2}}}\int_0^x {{y^{\frac{{\alpha \mu }}{2} - 1 - \frac{{{s_1}\alpha }}{2}}}{\rm{d}}y} {\rm{d}}{s_1},
\end{align}
where
\begin{equation}\label{afeaeg}
	\int_0^x {{y^{\frac{{\alpha \mu }}{2} - 1 - \frac{{{s_1}\alpha }}{2}}}{\rm{d}}y}   = \frac{{\Gamma \left( {\frac{{\alpha \mu }}{2} - \frac{{{s_1}\alpha }}{2}} \right)}}{{\Gamma \left( {\frac{{\alpha \mu }}{2} - \frac{{{s_1}\alpha }}{2} + 1} \right)}}{y^{\frac{{\alpha \mu }}{2} - \frac{{{s_1}\alpha }}{2}}}.
\end{equation}
Substituting~\eqref{afeaeg} into~\eqref{faekj} and using \cite[eq. (A-1)]{mathai2009h}, we derive the CDF as~\eqref{CDFfinal2}.

\section{Proof for Theorem~\ref{Theorem1}}\label{A2}
\renewcommand{\theequation}{C-\arabic{equation}}
\setcounter{equation}{0}
Substituting \eqref{PDFfinal} into \eqref{fdeklahjf}, we can express the achievable rate as
\begin{align}\label{ccguo}
{C_{\rm{k}}} =& \frac{{{\Gamma ^{ - 1}}\left( \mu  \right)\alpha }}{{2\Gamma \left( N \right)}}{\left( {\frac{{\text{1}}}{{{\text{2}}\pi i}}} \right)^2}\int_{{\mathcal{L}_1}} {\int_{{\mathcal{L}_2}} {\frac{{\Gamma \left( {{t_1} + \mu } \right)\Gamma \left( {{t_2}} \right)\Gamma \left( {N - {t_2}} \right)}}{{\Gamma \left( {\frac{\alpha }{2}{t_1} - {t_2}} \right)\Gamma \left( {\frac{\alpha }{2}{t_1}} \right)}}} } 
\notag\\&\times
I_4{\left( {\frac{\varphi }{{\beta {P_j}D_{jk}^{ - {\beta _k}}}}} \right)^{\frac{{ - \alpha }}{2}{t_1}}}{\left( {\frac{\varphi }{{{\eta _k}{P_I}}}} \right)^{{t_2}}}{\rm{d}}{t_1}{\rm{d}}{t_2}
\end{align}
where 
\begin{equation}
{I_4} = \int_0^\infty  {{y^{\frac{{ - \alpha }}{2}{t_1} - 1}}\log_2 \left( {1 + y} \right){\rm{d}}y}.
\end{equation}
With the help of \cite[eq. (2.6.9.21)]{Prudnikov1986Integrals} and \cite[eq. (8.334.3)]{gradshteyn2007}, $I_{4}$ can be deduced as
\begin{equation}
{I_4} = \frac{{\pi \csc \left( {\pi \frac{\alpha }{2}{t_1}} \right)}}{{\frac{\alpha }{2}{t_1}\ln 2}} = \frac{{\Gamma \left( {1 - \frac{\alpha }{2}{t_1}} \right){\Gamma ^{\text{2}}}\left( {\frac{\alpha }{2}{t_1}} \right)}}{{\Gamma \left( {{\text{1 + }}\frac{\alpha }{2}{t_1}} \right)\ln 2}}
\end{equation}
Substituting $I_4$ into \eqref{ccguo} and using \cite[eq. (A-1)]{mathai2009h}, we can derive the achievable rate as \eqref{ccfinal} to complete the proof.

\section{Proof for Theorem~\ref{Theorem2}}\label{A3}
\renewcommand{\theequation}{D-\arabic{equation}}
\setcounter{equation}{0}
Using the definition of Gamma function \cite[eq. (8.350)]{gradshteyn2007}, we can re-write $E_k$ as
\begin{equation}\label{ekudef}
E_k = \frac{{{\tau _1}^{{\tau _2}}}}{{2\Gamma \left( {{\tau _2}} \right)}}\int_0^\infty  {{x^{{\tau _2} - 1}}{e^{ - {\tau _1}x}}{F_{{\gamma _{k}}}}\left( x \right){\rm{d}}x}.
\end{equation}
Substituting \eqref{CDFfinal} into \eqref{ekudef}, we obtain
\begin{align}\label{faeg3}
&	{E_k} = \int_{{\mathcal{L}_1}} {\int_{{\mathcal{L}_2}} {\frac{{\Gamma \left( {{t_1} + \mu } \right)\Gamma \left( { - \frac{\alpha }{2}{t_1}} \right)\Gamma \left( {{t_2}} \right)\Gamma \left( {N - {t_2}} \right)}}{{\Gamma \left( {\frac{\alpha }{2}{t_1} - {t_2}} \right)\Gamma \left( {\frac{\alpha }{2}{t_1}} \right)\Gamma \left( {{\text{1}} - \frac{\alpha }{2}{t_1}} \right)}}} } {I_5}
\notag\\&\times\!\!\!
\frac{{\alpha {\tau _1}^{{\tau _2}}{\Gamma ^{ - 1}}\left( {{\tau _2}} \right)}}{{4\Gamma\!\left( N \right)\Gamma\!\left( \mu  \right)}}{\left(\! {\frac{{\text{1}}}{{{\text{2}}\pi i}}} \!\right)^2}{\left(\! {\frac{{D_{jk}^{{\beta _k}}\varphi }}{{\beta {P_j}}}} \!\right)^{\frac{{ - \alpha }}{2}{t_1}}}{\left(\! {\frac{\varphi }{{{\eta _k}{P_I}}}} \!\right)^{{t_2}}}{\rm{d}}{t_1}{\rm{d}}{t_2}
\end{align}
where
\begin{equation}
{I_5} = \int_0^\infty  {{x^{\frac{{ - \alpha }}{2}{t_1} + {\tau _2} - 1}}{e^{ - {\tau _1}x}}{\text{d}}x}.
\end{equation}
With the help of~\cite[eq. (3.351.3)]{gradshteyn2007} and~\cite[eq. (8.339.1)]{gradshteyn2007}, $I_D$ can be solved as 
\begin{equation}\label{id}
{I_5} = {\tau _1}^{\frac{\alpha }{2}{t_1} - {\tau _2}}\Gamma \left( {{\tau _2} - \frac{\alpha }{2}{t_1}} \right).
\end{equation}
Substituting $I_5$ into \eqref{faeg3} and using~\cite[eq. (A-1)]{mathai2009h}, we can derive \ref{berfinal} to complete the proof.

\end{appendices}
\bibliographystyle{IEEEtran}
\bibliography{Ref}

\end{document}